\documentclass[hyper]{JHEP} % 10pt is ignored!

\usepackage{epsfig,bm}
\newcommand\fverb{\setbox\pippobox=\hbox\bgroup\verb}
\newcommand\fverbdo{\egroup\medskip\noindent%
            \fbox{\unhbox\pippobox}\ }
\newcommand\fverbit{\egroup\item[\fbox{\unhbox\pippobox}]}
\newbox\pippobox

\title{$U(1)$ Invariant  $F(\tilde{R})$  Ho\v{r}ava-Lifshitz Gravity}

\author{J. Kluso\v{n}\\
Department of Theoretical Physics and Astrophysics \\
Faculty of Science, Masaryk University \\
Kotl\'{a}\v{r}sk\'{a} 2, 611 37, Brno \\
Czech Republic \\
E-mail: \email{klu@physics.muni.cz}}
\author{S. Nojiri \\ Department of Physics,  and\\
Kobayashi-Maskawa Institute for the Origin of Particles
and the Universe, \\ Nagoya University \\
 Nagoya 464-8602, Japan}
\author{S. D. Odintsov \\
Instituto de Ciencias del Espacio,
ICE(CSIC-IEEC), and\\
Instituci\`o Catalana de Recerca i Estudis Avan\c{c}ats (ICREA) \\
Campus UAB, Facultat de Ci\`encies, Torre C5-Parell-2a pl, \\ E-08193
Bellaterra (Barcelona) Spain; also at TSPU, Tomsk}
\author{D. S\'aez-G\'omez \\
Instituto de Ciencias del Espacio,
ICE(CSIC-IEEC), \\
Campus UAB, Facultat de Ci\`encies, Torre C5-Parell-2a pl, \\ E-08193
Bellaterra (Barcelona) Spain}
\preprint{}

\abstract{This paper is devoted to the study of various aspects of projectable
$F(R)$ Ho\v{r}ava-Lifshitz (HL) gravity.
We show that some versions of $F(R)$ HL gravity may have stable de Sitter
solution and unstable flat space solution.
In this case, the problem of scalar graviton does not appear because
flat space is not vacuum state.
Generalizing the $U(1)$ HL theory proposed in arXiv:1007.2410, we
formulate $U(1)$ extension of scalar theory and of $F(R)$
Ho\v{r}ava-Lifshitz gravity.
The Hamiltonian approach for such the theory is developed in full detail.
It is demonstrated that its Hamiltonian structure is the same as for the
non-relativistic covariant HL gravity.
The spectrum analysis performed around flat background indicates towards the
consistency of the theory because it contains graviton with only transverse
polarization.
Finally, we analyze the spatially-flat FRW equations for $U(1)$
invariant $F(R)$ Ho\v{r}ava-Lifshitz gravity.}


\def\be{\begin{equation}}
\def\ee{\end{equation}}
\def\bea{\begin{eqnarray}}
\def\eea{\end{eqnarray}}

\def\nn{\nonumber \\}
\def\e{\mathrm{e}}

\def\hN{\bar{N}}
\def\hK{\bar{K}}
\def\hnabla{\hat{\nabla}}

\def\mH{\mathcal{H}}

\def\tR{\tilde{R}}

\def\bx{\bm{x}}
\def\by{\bm{y}}

\newcommand{\mA}{\mathcal{A}}

\newcommand{\mG}{\mathcal{G}}

\def\mV{\mathcal{V}}

\newcommand{\bT}{\bm{T}}

\def\pb #1{\left\{#1\right\}}

\begin{document}

%%%%%%%%%%%%%%%%%%%%%
%%%%Introduction %%%%%%%%%
%%%%%%%%%%%%%%%%%%%%
\section{Introduction}\label{first}

In 2009 Petr Ho\v{r}ava formulated new proposal of quantum theory of gravity
that is power counting renormalizable \cite{Horava:2009uw}.
%,Horava:2008ih,Horava:2008jf}.
This theory is now known as Ho\v{r}ava-Lifshitz gravity (HL gravity).
It was also expected that this theory reduces to General
Relativity in the infrared (IR) limit.
The HL gravity is a new and intriguing formulation
of gravity as a theory with reduced
amount of symmetries and this fact leads to remarkable new phenomena
\footnote{For review and extensive list of references, see
\cite{Padilla:2010ge}.}.

The HL gravity is based on an idea that the Lorentz symmetry is restored in IR
limit of given theory and can be absent at high energy regime of given theory.
Explicitly,  Ho\v{r}ava considered systems whose scaling at short
distances exhibits a strong anisotropy between space and time,
\begin{equation}
\bx' =l \bx \, , \quad t' =l^{z} t \, .
\end{equation}
In $(D+1)$ dimensional space-time in order to have power counting
renormalizable theory  requires that $z\geq D$. It turns out however that
the symmetry group of given theory is reduced from the full diffeomorphism
invariance of General Relativity  to the foliation preserving diffeomorphism
\begin{equation}
\label{fpdi}
x'^i=x^i+\zeta^ i(t,\bx) \ , \quad t'=t+f(t) \, .
\end{equation}
The common  property of all modified theories of gravity is that whenever
the group of symmetries is restricted (as for example \ref{fpdi})
one more degree of freedom appears that is a spin-0 graviton.
An existence of this mode could be dangerous for all these
theories (for review, see \cite{Rubakov:2008nh}).
For example, in order to have the theory compatible
with observations one has to demand
that this scalar mode decouples in the IR regime.
Unfortunately, it seems that this might not be the case.
It was shown that the spin-0 mode is not stable in
the original version of the HL theory \cite{Horava:2009uw}
as well as in  SVW generalization \cite{Sotiriou:2009bx}.
Note that in both of these two versions, it was all
assumed the projectability condition that means that the lapse function $N$
depends on $t$ only. This assumption has a fundamental consequence for the
formulation of the theory since there is no local form of the Hamiltonian
constraint but the only global one.
Even if these instabilities indicate to problems with the projectable version
of the HL theory  it turns out that this is not the end of the whole story.
Explicitly, these instabilities are all found around the Minkowski background.
Recently, it was indicated that the de Sitter space-time is stable in the SVW
setup \cite{Huang:2010rq} and hence it seems to be reasonable to consider de
Sitter background as the natural vacuum of projectable version of the HL gravity.
This may be especially important for the theories with unstable flat space
solution.

On the other hand there is the second version of the HL gravity where the
projectability condition is not imposed
so that $N=N(\bx,t)$. Properties of
such theory were extensively studied in
\cite{Blas:2009yd}.
It was shown recently in \cite{Blas:2010hb} that  so called
healthy extended version of such theory could really be an interesting
candidate for the quantum theory of reality without ghosts and without
strong coupling problem despite its unusual Hamiltonian structure
\cite{Kluson:2010xx}.
Nevertheless, such theory is not free from its own internal problems.

Recently Ho\v{r}ava and Melby-Thompson \cite{Horava:2010zj}
proposed very interesting way to eliminate the spin-0 graviton.
They considered the projectable version of the HL gravity together
with extension of the foliation preserving diffeomorphism
to include a local $U(1)$ symmetry.
The resulting theory is then called as non-relativistic  covariant theory of
gravity \footnote{This theory was also studied in
\cite{Greenwald:2010fp,Huang:2010ay,daSilva:2010bm,
Kluson:2010zn}.}.
It was argued there \cite{Horava:2010zj} that the presence
of this new symmetry forces the coupling constant $\lambda$ to be equal
to one.
However, this result was questioned in \cite{daSilva:2010bm}
(see also \cite{Huang:2010ay,Kluson:2010zn})
where an alternative formulation of non-relativistic general covariant
theory of gravity was presented.
Furthermore, it was shown in \cite{Horava:2010zj,daSilva:2010bm}
that the presence of this new symmetry implies that the spin-0 graviton
becomes non-propagating and the spectrum of the linear fluctuations
around the background solution coincides with the fluctuation spectrum
of General Relativity.
This  construction was also extended to the case of RFDiff invariant
HL gravities \cite{Blas:2010hb,Kluson:2010na} in \cite{Kluson:2010zn}
where it was shown that the number of physical degrees of
freedom coincides with the number of physical degrees of freedom in General
Relativity.

The goal of this paper is to extend above construction to the case of
$F(\tR)$ HL gravities. These models were developed in series of papers
\cite{Chaichian:2010yi} \footnote{For further study in given direction, see
\cite{Kluson:2010xx,Chaichian:2010zn},
and for review, see \cite{Nojiri:2010wj}.}.
$F(\tR)$ HL gravity can be considered as natural
generalization of covariant $F(R)$ gravity.  Current interest to $F(R)$
gravity is caused by several important reasons.
First of all, it is known that such theory may give the unified
description of the early-time inflation and late-time acceleration
(for a review, see \cite{Nojiri:2010wj,Nojiri:2006ri}.)
Moreover, the whole sequence of the universe evolution
epochs: Inflation, radiation/matter dominance and dark energy may be
obtained within such theory.
The remaining freedom in the choice of $F(R)$ function could be used for
fitting the theory with observational data.
Second, it is known that higher derivatives gravity (like
$R^2$-gravity, for a review, see \cite{Buchbinder:1992rb}) has better
ultraviolet behavior than conventional General Relativity. Third, modified
gravity is pretending also to be the gravitational alternative for Dark
Matter.
Fourth, it is expected that  consistent quantum gravity emerging
from string/M-theory should be different from General Relativity.
Hence, it should be modified by fundamental theory.
Of course, all these reasons remain to be the same also for the HL gravity.
Additionally, it is expected that such modification may
be helpful for resolution of internal inconsistency problems of the HL theory.
Indeed, we will present the example of $F(\tR)$ HL gravity which has
stable de Sitter solution but unstable flat space
solution. In such a case, the original scalar graviton problem formally
disappears because flat space is not vacuum state. Hence, there is no sense
to study propagators structure around flat space.
The complete propagators structure should be investigated around
de Sitter solution which seems to be the candidate for vacuum space.

The paper is organized as follows.
In the next section, we briefly review the construction of $F(\tilde{R})$ HL
gravity. Section \ref{third} is devoted to study the de Sitter solutions and flat
space solutions in such theory.
Their stability is analyzed and it is shown that some versions of the theory may
have stable de Sitter but unstable flat space solution. Clearly, this is
indication that for such theories the appearance of scalar graviton is not a
problem due to fact that flat space is not vacuum solution. The whole spectrum
analysis should be developed around de Sitter vacuum. $U(1)$ extension of
$F(\tR)$ HL gravity as well as of scalar HL theory is given in section \ref{fourth}.
To check the consistency of such construction, two alternative
approaches to such extension are proposed.
The Hamiltonian structure of $U(1)$ invariant $F(\bar{R})$ HL
gravity is carefully investigated in section \ref{fifth}.
It is demonstrated that its Hamiltonian structure is the same
as in non-relativistic covariant HL gravity.
We also argue on the general grounds of the Hamiltonian formalism of
constrained system that the number of physical degrees of freedom coincides
with the number of physical degrees of freedom in $F(R)$ gravity despite of
the absence of the local Hamiltonian constraint. It is shown that for the
special case $F(x)=x$ the Hamiltonian structure of $U(1)$ invariant
$F(\bar{R})$ gravity coincides with the Hamiltonian structure of
non-relativistic covariant HL gravity found in \cite{Kluson:2010zn} which
can be considered as a nice check of our analysis.
The fluctuations around flat background for
$U(1)$ invariant  $F(R)$ are studied in section \ref{sixth}.
It is explicitly demonstrated that perturbations
spectrum contains the graviton with only transverse polarization. This
indicates that scalar graviton problem may be solved within such theory.
Section \ref{seventh} is devoted to demonstration that spatially-flat FRW
solutions of $U(1)$ invariant $F(\bar{R})$ gravity coincide with the
ones for the same theory without $U(1)$ invariance. Some summary and outlook
are given in last section.

%%%%%%%%%%%%%%%%%%%%%%%%%%%%%%%%%%%%%%%%%%%%%%%%%%%5
\section{Brief Review of $F(\tR)$ HL Gravity \label{second}}

In this section we give a brief review of $F(\tilde{R})$ HL gravity (for more
extensive review, see \cite{Nojiri:2010wj}).
This theory is naturally formulated in ADM formulation
of gravity \cite{gravitation}.
Let us consider $D+1$ dimensional manifold $\mathcal{M}$ with the
coordinates $x^\mu$, $\left(\mu=0,\cdots,D\right)$
and where $x^\mu=(t,\bx)$, $\bx=(x^1,\cdots,x^D)$.
We assume that this space-time is endowed with the metric
$\hat{g}_{\mu\nu}(x^\rho)$ with signature $(-,+,\cdots,+)$.
Suppose that $\mathcal{M}$ can be foliated by a family of space-like surfaces
$\Sigma_t$ defined by $t=x^0$.
Let $g_{ij}$, $\left(i,j=1,\cdots,D\right)$ denotes the
metric on $\Sigma_t$ with inverse $g^{ij}$ so that
$g_{ij}g^{jk}= \delta_i^k$.
We further introduce the operator $\nabla_i$ that is covariant derivative
defined with the metric $g_{ij}$.
We introduce the future-pointing unit normal vector $n^\mu$ to the surface $\Sigma_t$.
In ADM variables one has $n^0=\sqrt{-\hat{g}^{00}}$,
$n^i=-\hat{g}^{0i}/\sqrt{-\hat{g}^{00}}$.
We also define  the lapse function $N=1/\sqrt{-\hat{g}^{00}}$ and
the shift function $N^i=-\hat{g}^{0i}/\hat{g}^{00}$.
In terms of these variables  the
components of the metric $\hat{g}_{\mu\nu}$ are written as
\begin{eqnarray}
&& \hat{g}_{00}=-N^2+N_i g^{ij}N_j \, , \quad \hat{g}_{0i}=N_i \, ,
\quad \hat{g}_{ij}=g_{ij} \, , \nonumber \\
&& \hat{g}^{00}=-\frac{1}{N^2} \, , \quad \hat{g}^{0i}=\frac{N^i}{N^2} \, ,
\quad \hat{g}^{ij}=g^{ij}-\frac{N^i N^j}{N^2}\, .
\nonumber %\\
\end{eqnarray}
Then it is easy to see that
\begin{equation}
\sqrt{-\det \hat{g}}=N\sqrt{\det g} \, .
\end{equation}
The extrinsic derivative is defined as
\begin{equation}
K_{ij}=\frac{1}{2N} (\partial_t g_{ij}-\nabla_i N_j- \nabla_j N_i) \, .
\end{equation}
It is well-known that the components of
the Riemann tensor can be written in terms of ADM variables.
For example, in case of Riemann curvature we have
\begin{equation}
\label{R}
{}^{(D+1)}R=K^{ij}K_{ij}-K^2+R+\frac{2}{\sqrt{-\hat{g}}}
\partial_\mu\left(\sqrt{-\hat{g}}n^\mu K\right)
 - \frac{2}{\sqrt{g}N}\partial_i
\left(\sqrt{g}g^{ij}\partial_j N\right) \ ,
\end{equation}
where $R$ is $D-$dimensional curvature.
The general formulation of Ho\v{r}ava-Lifshitz $F(\tR)$ gravity
was presented in series of papers in \cite{Chaichian:2010yi}
\footnote{For further study in given direction, see
\cite{Kluson:2010xx,Chaichian:2010zn},
and for review, see \cite{Nojiri:2010wj}}.
This construction is based on the modification of the relation (\ref{R}).
In fact, the action introduced in \cite{Chaichian:2010yi} takes the form
\begin{equation}
\label{actionNOJI}
S_{F(\tilde{R})}= \frac{1}{\kappa^2} \int dt d^D\bx
\sqrt{g}N F (\tilde{R}) \, ,
\end{equation}
where
\begin{equation}
\label{tR} \tilde{R}= K_{ij}\mG^{ijkl}K_{kl}
+ \frac{2\mu}{\sqrt{-\hat{g}}} \partial_\mu \left(\sqrt{-\hat{g}}n^\mu K\right)
 -\frac{2\mu}{\sqrt{g}N} \partial_i \left(\sqrt{g}g^{ij}\partial_j N\right)
 -\mathcal{V}(g) \, ,
\end{equation}
where $\mu$ is constant, $K=K_{ij}g^{ji}$ and where  the
generalized de Witt metric $\mG^{ijkl}$  is defined as
\begin{equation}
\mG^{ijkl}=\frac{1}{2}(g^{ik}g^{jl} + g^{il}g^{jk})-\lambda g^{ij}g^{kl} \, ,
\end{equation}
where $\lambda$ is real constant.
Finally $\mV(g)$ is general function of $g_{ij}$ and its covariant derivatives.
We should also note that the special case of $F(\hat{R})$ HL gravity where
$\mu=0$ was introduced in \cite{Kluson:2009xx}.
An important drawback of the case $\mu=0$ is that it
cannot lead  to the FRW cosmological equations directly.
The FRW equations may be obtained only as the limit
$\mu=0$ from general theory.
In fact, due to the absence of the derivative
terms in (\ref{tR})  this theory cannot flow to $F(R)$ gravity action in IR.

We conclude this section with the remark that the action (\ref{actionNOJI}) is
invariant under restricted group of symmetries with is foliation
preserving diffeomorphism
\begin{equation}
\label{fpd}
t'-t=f(t) \, , \quad x'^i-x^i=\xi^i(t,\bx) \, .
\end{equation}
Note that under this transformations the metric components  transform as
\begin{eqnarray}
\label{trmet}
N'_i(\bx',t')&=&N_i(\bx,t)-N_i( \bx,t)\dot{f}-N_j(\bx,t)\partial_i\xi^j(\bx,t)
-g_{ij}(\bx,t)\dot{\xi}^j(\bx,t) \, ,
\nonumber \\
N'(t')&=&N(t)-N(t)\dot{f} \, , \nonumber \\
g'_{ij}(\bx',t')&=& g_{ij}(\bx,t) - g_{ik}(\bx,t)\partial_j\xi^k(\bx,t)
 - \partial_i\xi^k(\bx,t)g_{kj}(\bx,t) \, .
\nonumber %\\
\end{eqnarray}
%%%%%%%%%%%%%%%%%%%%%%%%%%%%%%%%%%%%%

\section{Solutions and their stability in $F(\tilde{R})$ HL gravity \label{third}}

In this section, we consider different solutions  of standard $F(\tilde{R})$
Ho\v{r}ava-Lifshitz gravity, specially de Sitter and vacuum solutions. The
stability of this kind of solutions is studied, and it is shown that it
depends completely on the choice of function $F(\tilde{R})$.
This suggests the way to resolve the scalar graviton
problem of Ho\v{r}ava-Lifshitz gravity:
Its $F(\tilde{R})$ version may have stable de Sitter vacuum but not
flat-space which turns out to be unstable. In this situation, the
problem is solved simply due to the fact that space flat is not vacuum. In
order to study the consistency of the theory its spectrum in de Sitter space
should be investigated what lies beyond the scopes of this work.

%%%%%%%%%%%%%%%%%%%%%%%%%%%%%%%%%%%%%%%%%%%%%%%%%
\subsection{Stability of de Sitter solutions
in $F(\tilde{R})$ gravity}

Let us consider  the stability of the de Sitter solution. As dark energy and
even inflation may be described (in their simplest form) by the de Sitter
space, its stability becomes very important topic. Especially in the case
of inflation, where a graceful exit is needed to enter to radiation/matter
dominance, de Sitter space should be unstable.
In general, standard $F(R)$ gravity contains several de Sitter
points, which represent critical points (see \cite{FRdeSitter}).
This analysis can be extended to $F(\tilde{R})$ Ho\v{r}ava-Lifshitz gravity.
Let us write the first FRW equation in $F(\tilde{R})$ Ho\v{r}ava-Lifshitz
gravity \cite{Chaichian:2010yi}
\be
0=F(\tilde{R})-6\left[(1-3\lambda
+3\mu)H^2+\mu\dot{H}\right]F'(\tilde{R})
+6\mu H \dot{\tilde{R}}F''(\tilde{R})-\kappa^2\rho_m-\frac{C}{a^3}
\, ,
\label{1.18}
\ee
For a given
$F(\tilde{R})$, de Sitter solution $H(t)=H_0$, where $H_0$ being a
constant, has to satisfy the first equation FRW equation (\ref{1.18}),
\be
0=F(\tilde{R}_0)-6H^2_0(1-3\lambda+3\mu)F'(\tilde{R}_0)\, ,
\label{2.1}
\ee
where $C=0$ and  it is assumed the absence of any kind of
matter. The scalar $\tilde{R}$ is given in this case by,
\be
\tilde{R}_0=3(1-3\lambda+6\mu)H_0^2\, .
\label{2.2}
\ee
Then, the positive roots of equation (\ref{2.1}) are the
de Sitter points allowed by a particular choice of an $F(\tilde{R})$
function. Assuming de Sitter solution, one can write $F(\tilde{R})$ around
$\tilde{R}_0$ as a series,
\be
F(\tilde{R})=F_0+F'_0(\tilde{R}-\tilde{R}_0)
+\frac{F''_0}{2}(\tilde{R}-\tilde{R}_0)^2
+\frac{F^{(3)}_0}{6}(\tilde{R}-\tilde{R}_0)^3+O(\tilde{R}^4)\, .
\label{2.3}
\ee
Here, the primes denote derivative with respect to
$\tilde{R}$ while the subscript $0$ means that it is evaluated in
$\tilde{R}_0$.
Then, we can perturb the solution writing the Hubble parameter as,
\be
H(t)=H_0+\delta(t)\, .
\label{2.4}
\ee
Using the function $F(\tilde{R})$ evaluated around a given
de Sitter solution (\ref{2.3}), and the perturbed solution (\ref{2.4}) in the
first FRW equation (\ref{1.18}), it yields,
\bea
0 &=& \frac{1}{2}F_0-3H_0^2(1-3\lambda +3\mu) \nonumber \\
&& -3H_0\left[\left((1-3\lambda)F_0'
+6F_0''H_0^2(-1+3\lambda-6\mu)(-1+3\lambda-3\mu)\right)\delta(t)
\right. \nonumber \\
&& \left.+6F_0''\mu
H_0(-1+3\lambda-3\mu)\dot{\delta}(t)-12F_0''\mu^2\ddot{\delta}(t)\right]\, .
\label{2.5}
\eea
Here, we have taken the linear approach on  $\delta$ and
its derivatives. Note that the first two terms in the equation (\ref{2.5})
can be dropped because of the equation (\ref{2.1}), which is assumed to be
satisfied.
Then, equation (\ref{2.5}) can be written in a more convenient way as,
\bea
&& \ddot{\delta}(t)+\frac{H_0(1-3\lambda+9\mu)}{2\mu}\dot{\delta}(t) \nn
&& +\frac{1}{12\mu^2}\left[(3\lambda-1)\frac{F_0'}{F_0''}
 -6H_0^2(1-3\lambda+6\mu)(1-3\lambda+3\mu)\right]\delta(t)=0\, .
\label{2.6}
\eea
Then, the perturbations of de Sitter solution will depend completely on the
model,that is on the derivatives of the $F(\tilde{R})$ function, as well as on
the parameters of the theory $(\lambda,\mu)$. Note that the
instability will be large if the term in front of $\delta(t)$ is negative, as
the perturbations will increase exponentially, while if we have a
positive frequency, the perturbations will behave as a damped harmonic
oscillator. During dark energy epoch, as the scalar curvature is very small,
the IR limit of the theory can be assumed, where General Relativity  is recovered,
and in such a case we have $\lambda=\mu\sim 1$, and the frequency will depend
completely on the value of $\frac{F_0'}{F_0''}$.
In order to avoid large instabilities during the dark energy phase, the condition
$\frac{F_0'}{F_0''}>12H_0^2$ has to be imposed. Nevertheless, during the
inflationary epoch, where the scalar curvature is large, the IR limit is not
a convenient approach, and the perturbations will depend also on the
values of $(\lambda,\mu)$.
Although if we assume a very small $F_0''$, the
first term in the frequency of the equation (\ref{2.6}) will dominate and
if $\lambda>1/3$, the stability of the solution will depend on the sign of
$\frac{F_0'}{F_0''}$, being stable when such a coefficient is positive.

\subsection{On flat space solution in $F(\tilde{R})$ gravity}

Let us now study flat space solutions in $F(\tilde{R})$ HL  gravity. In this
section we restrict to the case of $3+1$ dimensional space-time.  A general
metric in the  ADM decomposition in a $3+1$ space-time  is given by,
\be
ds^2=-N^2 dt^2+g^{(3)}_{ij}(dx^i+N^idt)(dx^j+N^jdt)\, ,
\label{2.7}
\ee
where $i,j=1,2,3$, $N$ is the so-called lapse variable,
and $N^i$ is the shift $3$-vector.
For flat space  the variables from the metric (\ref{2.7}) take the values,
\be
N=1\, , \quad N_i=0 \quad \mbox{and} \quad g_{ij}=g_0\delta_{ij}\, ,
\label{2.8}
\ee
where $g_0$ is a constant. Then, the scalar curvature
$\tilde{R}=0$, and so that our theory has flat space solution, the function
$F(\tilde{R})$  has to satisfy,
\be
F(0)=0\, .
\label{2.9}
\ee
Hence, we assume  the condition (\ref{2.9}) is satisfied, otherwise the theory
has no flat space solution.
We are interested to study the stability of such solutions for a general
$F(\tilde{R})$, by perturbing the metric in vacuum (\ref{2.8}), this yields,
\be
N=1+\delta_N(t) \quad \mathrm{and} \quad
g_{ij}=g_0(1+\delta_g(t))\delta_{ij}\, .
\label{2.10}
\ee
For simplicity, we restrict the study on the time-dependent perturbations (no
spatial ones) and on the diagonal terms of the metric. Note that, as we are
assuming the projectability condition, by performing a transformation of
the time coordinate, we can always rewrite $N=1$.
Then, the study of the perturbations is focused on the spatial
components of the metric $ g_{ij}$, which can be written in a more
convenient way as,
\be
g_{ij}=g_0 \left(1+\int \delta(t)\right)\delta_{ij}
\sim g_0\e^{\int \delta(t)dt}\delta_{ij}\, .
\label{2.11}
\ee
By inserting (\ref{2.11}) in the first field equation, obtained by
the variation of the action on $N$, it yields at lowest order on $\delta(t)$,
\be
12\mu^2F_0''\ddot{\delta}(t)\delta(t)-6\mu^2F_0''\dot{\delta}^2(t)
+(-1+3 \lambda)F_0'\delta^2(t) =0\, .
\label{2.12}
\ee
Here the derivatives $F_0'$, $F_0''$ are evaluated on $\tilde{R}=0$.
Then, the perturbation $\delta$ will depend completely on the kind of theory
assumed. We can study some general cases by imposing conditions on the
derivatives of $F(\tilde{R})$.
\begin{itemize}
\item For the  case $F_0''=0$, it gives $\delta(t)=0$,
such that at lowest order the flat space is completely stable for this case.
\item For $F_0'=0$ and $F_0''\neq0$, the differential equation (\ref{2.12})
has the solution,
\be
\delta(t)=C_1t\left(1+\frac{C_1t}{C_2}\right)+C_2
\label{2.13}
\ee
where $C_{1,2}$ are integration constants.
Then, for this case, the perturbations grow as the power of the time coordinate, and flat
space becomes unstable.
\end{itemize}

Hence, depending on the theory, the flat space solution will be stable or
unstable, which becomes very important as it could be used to distinguish the
theories or analyze their consistency.

\subsection{A simple example}

Let us now discuss a simple example.
We consider the function,
\be
F(\tilde{R})=\kappa_0\tilde{R}+\kappa_1\tilde{R}^n\, ,
\label{2.14}
\ee
where $\kappa_{0,1}$ are coupling constants and $n>1$.
Note that this family of theories satisfies the condition (\ref{2.9}).
The values of first and second derivatives evaluated in the solution depend on
the value of $n$ in (\ref{2.14}),
\be
F'(0)=\kappa_0\, , \quad
F''(0)=\kappa_1n(n-1)\tilde{R}^{n-2}\, .
\label{2.15}
\ee
Then, we can distinguish between the cases,
\begin{itemize}
\item For $n\neq2$, we have $F''(0)=0$,
and by the analysis performed above, it follows that flat space is stable
\item For $n=2$, we have $F''(0)=\kappa_1$, and flat space is unstable.
\end{itemize}
Hence, we have shown  that the stability of solution
depends completely on the details of the theory.

We can now analyze the de Sitter solution.
Using the equation (\ref{2.1}), we can find the de Sitter
points allowed by the class of theories given in Eq.~(\ref{2.14}),
\be
\frac{3}{2}H_0^2(3\lambda-1)\kappa_0
+\frac{\kappa_1\left(3H_0^2(1-3\lambda+6\mu)\right)^n(1-3\lambda
+6\mu-2n(1-3\lambda+3\mu))}{2(1-3\lambda+6\mu)}=0\, .
\label{2.16}
\ee
Resolving the Eq.~(\ref{2.16}),  de Sitter solutions are obtained.
For simplicity, let us consider  $n=2$, in such a case the
equation $(\ref{2.16})$ has two roots for $H_0$ given by,
\be
H_0= \pm
\frac{\sqrt{\kappa_0(3\lambda-1)}}{3\sqrt{\kappa_1(1-3\lambda+6\mu)(-1+3\lambda-2\mu)}}
\, .
\label{2.17}
\ee
As we are interested in de Sitter points, we just consider the positive root in
(\ref{2.17}). Then, the stability of such de Sitter point can be analyzed by
studying the derivatives of the function $F(\tilde{R})$ evaluated in $H_0$.
The stability will depend on the value of $\frac{F_0'}{F_0''}$, which
for this case yields,
\be
\frac{F_0'}{F_0''}=\frac{\kappa_0}{2\kappa_1} +12H_0^2\, .
\label{2.18}
\ee
In the IR limit of the theory ($\lambda\rightarrow1$,
$\mu\rightarrow1$), the condition for the stability of de Sitter points
$\frac{F_0'}{F_0''}>12H_0^2$  is clearly satisfied by (\ref{2.18}).
Even in the non IR limit, $F_0''=2\kappa_1$
and assuming $\kappa_1\ll 1$, we have that,
\be
\frac{F_0'}{F_0''}=3H_0^2(1-3\lambda+6\mu)
+\frac{\kappa_0}{2\kappa_1}\, .
\label{2.19}
\ee
As this term is positive, we have that the instabilities will oscillate and be
damped, such that the de Sitter point becomes stable.

Thus, we presented the example of the $F(\tilde{R})$ theory where flat space
is unstable solution and de Sitter space is stable solution. The problem of
scalar graviton does not appear in this theory because one has to analyze
the spectrum of theory around de Sitter space which is real vacuum.
Indeed, flat space is not stable and cannot be considered as the vacuum
solution.
Of course, deeper analysis of de Sitter spectrum structure of the theory is
necessary.
Nevertheless, as we see already standard $F(\tilde{R})$ gravity suggests
the way to resolve the pathologies which are well-known in
Ho\v{r}ava-Lifshitz gravity.

%%%%%%%%%%%%%%%%%%%%%%%%%%%%%%%%%%%%%%%%%%%%%%%%%%%55
\section{$U(1)$ Invariant $F(\bar{R})$ Ho\v{r}ava-Lifshitz Gravity \label{fourth}}

Our goal is to see whether it is possible to extend the gauge symmetries
for above action as in \cite{Horava:2010zj}.
As the first step we introduce two non-dynamical fields $A,B$ and
rewrite the action (\ref{actionNOJI}) into the form
\begin{equation}
S_{F(\tilde{R})}= \frac{1}{\kappa^2}\int dt d^D\bx
\sqrt{g}N (B(\tilde{R}-A)+F(A)) \, .
\end{equation}
It is easy to see that solving the equation of motion with respect to
$A,B$ this action reduces into (\ref{actionNOJI}). On the other hand
when we perform integration by parts we obtain the action in the form
\begin{eqnarray}
\label{SFtR}
S_{F(\tilde{R})} =\frac{1}{\kappa^2}\int dt d^D\bx \left( \sqrt{g}N B(
K_{ij}\mG^{ijkl}K_{kl} -\mV(g)-A) %+
\nonumber \right. \\
\left. +\sqrt{g}N F(A) -2\mu \sqrt{g}N\nabla_n B
K  + 2\mu \partial_i B \sqrt{g}g^{ij}
\partial_j N \right) \, , %\nonumber %\\
\end{eqnarray}
where we ignored the boundary terms and where
\begin{equation}
\nabla_n B=\frac{1}{N} (\partial_t B-N^i\partial_i B) \, .
\end{equation}
Let us now introduce $U(1)$ symmetry where the shift function transforms as
\begin{equation}
\label{deltaNi}
\delta_\alpha N_i(\bx,t)= N(\bx,t)\nabla_i\alpha(\bx,t) \, .
\end{equation}
It is important to stress that as opposite to the case of pure
Ho\v{r}ava-Lifshitz gravity the kinetic term is multiplied with $B$ that is
space-time dependent and hence it is not possible to perform similar
analysis as in \cite{Horava:2010zj}.
This procedure frequently uses the integration by parts and the fact that
covariant derivative annihilates metric tensor together with the crucial
assumption that $N$ depends on time only. Now due to the presence of $B$
field we have to  proceed step by step with the construction of the action
invariant under (\ref{deltaNi}).
As the first step note that under (\ref{deltaNi}) the kinetic term
$S^\mathrm{kin}=\frac{1}{\kappa^2}\int dt d^D\bx \sqrt{g}K_{ij}\mG^{ijkl}K_{kl}$
transforms as
\[
\delta_\alpha S^\mathrm{kin} =-\frac{2}{\kappa^2}\int dt
d^D\bx\sqrt{g}NB K_{ij}\mG^{ijkl}\nabla_i \nabla_j
\alpha \, .
%\nonumber %\\
\]
In order to compensate this variation of the action we introduce new
scalar field $\nu$ that under (\ref{deltaNi}) transforms as
\begin{equation}
\label{deltaNinu}
\delta_\alpha \nu(t,\bx)=\alpha(t,\bx)
\end{equation}
and add to the action following term
\begin{equation}
S_\nu^{(1)}= \frac{2}{\kappa^2}\int dt d^D\bx\sqrt{g}NB
K_{ij}\mG^{ijkl}\nabla_i \nabla_j \nu \, .
\nonumber %\\
\end{equation}
Note that under (\ref{deltaNinu}) this term transforms as
\[
\delta_\alpha S_\nu^{(1)}= \frac{2}{\kappa^2}\int dt
d^D\bx\sqrt{g}NB K_{ij}\mG^{ijkl}\nabla_k \nabla_l \alpha %-
%\nonumber \\
 - \frac{2}{\kappa^2}\int dt d^D\bx\sqrt{g}NB \nabla_i\nabla_j
\alpha \mG^{ijkl}\nabla_k \nabla_l \nu \, .
\]
%\end{eqnarray}
so that we add the second term into the action
\begin{equation}
S_{\nu}^{(2)}= \frac{1}{\kappa^2}\int dt d^D\bx \sqrt{g}N B
\nabla_i\nabla_j\nu \mG^{ijkl} \nabla_k\nabla_j \nu \, .
\end{equation}
As a result, we find that $S^\mathrm{kin}+S_\nu^{(1)}+ S_\nu^{(2)}$ is
invariant under (\ref{deltaNi}) and (\ref{deltaNinu}).

As the next step we analyze the variation of the $B$-kinetic part of
the action $S^{B\mathrm{kin}}=-\frac{2\mu}{\kappa^2}
\int dt d^D\bx  \sqrt{g} N \nabla_n B K$ under the variation (\ref{deltaNi})
\be
\label{alphaBn}
\delta_\alpha S^{B\mathrm{kin}}
=\frac{2\mu}{\kappa^2} \int dt d^D\bx
\sqrt{g}\alpha \nabla^i(\nabla_i B
K) +\frac{2\mu}{\kappa^2} \int dt d^D\bx
\sqrt{g}N \alpha\nabla_i\nabla_j(
g^{ij}\nabla_n B) \ .
%\nonumber %\\
\ee
We see that in order to cancel this variation it is appropriate to add
following expression into the action
\begin{eqnarray}
S^{\nu-B}&=& -\frac{2}{\kappa^2}\mu \int dt d^D\bx
\sqrt{g}N\nu\nabla^i(\nabla_i B K) -\frac{2}{\kappa^2}\mu\int dt d^D\bx
\sqrt{g}N \nu\nabla_i\nabla_j[ g^{ij}\nabla_n B]
\nonumber \\
&& +\frac{2}{\kappa^2}\mu \int dt d^D\bx \sqrt{g}N \nabla^k \nu \nabla_k
B \nabla_i\nabla^i\nu \, . \nonumber %\\
\end{eqnarray}
Then it is easy to see that $S^{B\mathrm{kin}}+S^{\nu-B}$ is invariant under
(\ref{deltaNi}) and (\ref{deltaNinu}).
Collecting all these results we find following $F(\hat{R})$ HL action that
is invariant under (\ref{deltaNi}) and (\ref{deltaNinu})
\begin{eqnarray}
\label{SFtR2}
S_{F(\tilde{R})} &=&\frac{1}{\kappa^2}\int dt d^D\bx \left( \sqrt{g}N B(
K_{ij}\mG^{ijkl}K_{kl} -\mV(g)-A)  \nonumber \right. \\
&& \left. +\sqrt{g}N F(A) -2\mu \sqrt{g}N\nabla_n B K + 2\mu
\partial_i B \sqrt{g}g^{ij} \partial_j N \right) \nonumber \\
&& - 2\mu \int d^D\bx dt \sqrt{g}\nu\nabla^i(\nabla_i B K)
-2\mu\int d^D\bx dt \sqrt{g}N \nu\nabla_i\nabla_j(
g^{ij}\nabla_n B)  \nonumber \\
&& + 2\mu \int dt d^D\bx \sqrt{g} N \nabla^i \nu \nabla_i B \nabla^
j\nabla_j\nu \nonumber \\
&& + 2\int d^D\bx\sqrt{g}NB K_{ij}\mG^{ijkl}\nabla_k \nabla_l \nu
+\int d^D\bx \sqrt{g}B \nabla_i\nabla_j\nu \mG^{ijkl}
\nabla_k\nabla_l \nu \, .
%\nonumber %\\
\end{eqnarray}
Note that we can write this action in suggestive form
\begin{eqnarray}
\label{SFtR3}
S_{F(\tilde{R})} &=& \frac{1}{\kappa^2}\int dt d^D\bx \left( \sqrt{g}N B(
(K_{ij}+\nabla_i\nabla_j\nu)\mG^{ijkl}(K_{kl}+\nabla_k\nabla_l \nu)
-\mV(g)-A) \nonumber \right. \\
&& \left. +\sqrt{g}N F(A) -2\mu
\sqrt{g}N(\nabla_n B+\nabla^i \nu\nabla_i B)
g^{ij}(K_{ji}+\nabla_j\nabla_i\nu) + 2\mu
\partial_i B \sqrt{g}g^{ij} \partial_j N \right)
\nonumber \\
\end{eqnarray}
or in even more suggestive form by introducing
\begin{equation}
\hN_i=N_i-N\nabla_i \nu  \, ,
\quad \hK_{ij}=\frac{1}{2N}(\partial_t g_{ij}
 -\nabla_i \hN_j-\nabla_j \hN_i)
\end{equation}
so that
\begin{eqnarray}
\label{SFtRfinal}
S_{F(\tilde{R})} &=& \frac{1}{\kappa^2}\int dt d^D\bx \left( \sqrt{g}N B(
\hK_{ij}\mG^{ijkl}\hK_{kl} -\mV(g)-A) \nonumber \right. \\
&& \left. +\sqrt{g}N F(A) -2\mu \sqrt{g}N\hnabla_n B
g^{ij}\hK_{ji}  + 2\mu \partial_i B \sqrt{g}g^{ij}
\partial_j N \right)
%\nonumber \\
\end{eqnarray}
is formally the same as the $F(\tR)$ Ho\v{r}ava-Lifshitz gravity action.
Clearly this action is invariant under arbitrary $\alpha=\alpha(t,\bx)$.
Moreover, such an introduction of $U(1)$ symmetry is trivial and does not
modify the physical properties of the theory.
This is nicely  seen from the fact that $\nu$ appears in the action in the
combination with $N_i$ through $\bar{N}_i$ where $\nu$ plays the role of
the St\"{u}ckelberg field. In order to get physical content of given symmetry
we follow \cite{Horava:2010zj} and \cite{daSilva:2010bm}
and introduce following  term into action
\begin{equation}
\label{Snuk}
S^{\nu,k} = \frac{1}{\kappa^2}\int dt d^D\bx  \sqrt{g}
B\mG(g_{ij})(\mA-a) \, ,
\end{equation}
where
\begin{equation}
a=\dot{\nu}-N^i\nabla_i\nu + \frac{N}{2}
\nabla^i \nabla_i \nu \, ,
\end{equation}
where $\dot{X}\equiv \frac{dX}{dt}$.
Note $a$ transforms under $\alpha$ variation as
\[
a'(t,\bx) = a(t,\bx)+\dot{\alpha}(t,\bx)-N^i(t,\bx)
\nabla_i\alpha(t,\bx) \, .  \nonumber %\\
\]
Then it is natural to suppose that
$\mA$ transforms under $\alpha$-variation as
\begin{equation}
\mA'(t,\bx)=\mA(t,\bx)+\dot{\alpha}(t,\bx)
 - N^i(t,\bx)\nabla_i \alpha(t,\bx)
\end{equation}
so that we immediately see that $\mA-a$ is invariant under $\alpha$-variation.
The function $\mG$ can generally depend on arbitrary combinations of metric $g$
and matter field and we only demand that it should
be invariant under foliation preserving diffeomorphism
(\ref{fpd}) and under (\ref{deltaNi}) and (\ref{deltaNinu}).
For our purposes it is, however, sufficient to restrict ourselves to the models
where $\mG$ depends on the spatial curvature $R$ only.
Now one observes that the equation of motion for $\mA$ implies the constraint
\begin{equation}
B\mG(R)=0
\end{equation}
that for non-zero $B$ implies the condition $\mG(R)=0$.
Note that this condition is crucial for elimination of the
scalar graviton when we study fluctuations around flat background. We demonstrate
this important result in the next section.

Finally note that it is possible to integrate out $B$ and $A$ fields from the
actions (\ref{SFtRfinal}) and (\ref{Snuk}) that leads to
\begin{eqnarray}
\label{SFbarRf}
S_{F(\bar{R})}&=& \frac{1}{\kappa^2} \int dt d^D\bx \sqrt{g}N
F(\bar{R}) \, , \nonumber \\
\bar{R}&=&\hK_{ij}\mG^{ijkl}\hK_{kl}-\mV(g)+
\frac{2\mu}{\sqrt{g} N} \left\{ \partial_t \left( \sqrt{g} \bar K \right)
 - \partial_i \left( \sqrt{g} N^i \bar{K} \right) \right\}+
\frac{1}{N}\mG(\mA-a) \,  .
\nonumber \\
\end{eqnarray}
This finishes the construction of $U(1)$ invariant  $F(\bar{R})$ HL
theory action.

%%%%%%%%%%%%%%%%%%%%%%%%%%%%%%%%%%%%%%%%%%%%%%%
\subsection{Lagrangian for the Scalar Field}

Now we extend above analysis to the action for the matter field with the
following general form of the scalar field action
\begin{equation}
\label{Smatt}
S_\mathrm{matt}=-\int dt d^D\bx \sqrt{g}N X \, ,
\end{equation}
where
\begin{equation}
X=-(\nabla_n\phi)^2 + F(g^{ij}\partial_i\phi\partial_j\phi) \, .
\end{equation}
where $F(x)=X+\sum_{n=2}^z X^n$ and where we defined
\begin{equation}
\nabla_n\phi=\frac{1}{N}(\partial_t\phi-N^i \partial_i \phi) \, .
\end{equation}
Note that this general  form of the scalar field action is consistent with
the anisotropy of target space-time as was shown in \cite{Kiritsis:2009sh}.

Now we try to extend above action in order to make it invariant under (\ref{deltaNi}).
Note that under this variation the scalar field  action (\ref{Smatt}) transforms as
\be
\label{deltaSmatt}
\delta_\alpha S_\mathrm{matt}= -2\int dt d^D\bx \sqrt{g}N
\nabla^i\alpha \nabla_i\phi \nabla_n\phi
%\nonumber %\\
\ee
using
\begin{equation}
\label{deltaX} \delta_\alpha X=
2\nabla^i\alpha\nabla_i\phi \nabla_n\phi \, .
\end{equation}
We compensate the variation (\ref{deltaSmatt}) by introducing
additional term into action
\be
\label{Smattnu}
S_{\mathrm{matt-}\nu} = -2\int dt d^D\bx \sqrt{g}\nu N \nabla^i
(\nabla_i\phi\nabla_n\phi)
%\nonumber \\ &&
+ \int dt d^D\bx \sqrt{g}
\nabla^i \nu \nabla^j \nu \nabla_i\phi \nabla_j\phi %\nonumber \\
\ee
Then the action (\ref{Smattnu}) transforms under
(\ref{deltaNi}) and (\ref{deltaNinu}) as
\begin{equation}
\delta_\alpha S_{\mathrm{matt-}\nu}=
2\int dt d^D\bx  \sqrt{g}N
\nabla^i \alpha \nabla_i\phi\nabla_n\phi
\end{equation}
that compensates the variation (\ref{deltaSmatt}).

In the same way one can analyze more general form of
the scalar action
\begin{equation}
\label{Smatgen}
S_\mathrm{matt}=-\int dt d^D\bx \sqrt{g}N K(X) \, .
\end{equation}
In order to find the generalization of the action (\ref{Smatgen})
which is invariant under (\ref{deltaNi}) we introduce two
auxiliary fields $C$, $D$ and write the action (\ref{Smatgen}) as
\begin{equation}
\label{Smattex}
S_\mathrm{matt}=-\int dt d^D\bx \sqrt{g}N [K(C)+D(X-C)] \, .
\end{equation}
Clearly this action transforms under (\ref{deltaNi}) as
\be
\label{deltaSmattG}
\delta_\alpha S_\mathrm{matt}= -\int dt d^D\bx
\sqrt{g}N D \delta_\alpha X= -2\int  dt d^D\bx
\sqrt{g}N D \nabla^i\alpha \nabla_i\phi \nabla_n\phi \, ,
%\nonumber \\
\ee
where relation (\ref{deltaX}) is used. It is easy to see that the
variation of the  following term
\be
\label{SmattnuG}
S_{\mathrm{matt}-\nu} = 2\int dt d^D\bx \sqrt{g} N D
\nabla^i\nu \nabla_i\phi\nabla_n\phi
+ % \nonumber \\ &+&
\int dt d^D\bx  \sqrt{g} D \nabla^i \nu \nabla^j \nu \nabla_i\phi
\nabla_j\phi %\nonumber \\
\ee
compensates the variation (\ref{deltaSmattG}).
Finally note that (\ref{Smattex}) together
with (\ref{SmattnuG}) can be written in more elegant form
\be
\label{Smattin}
S_\mathrm{matt} = -\int  dt d^D\bx  \sqrt{g}N
[K(C)+D(\bar{X}-C)]= -\int dt d^D\bx
\sqrt{g}N K(\bar{X}) \, , %\nonumber \\
\ee
where
\begin{eqnarray}
\bar{X}&=&-(\bar{\nabla}_n\phi)^2+
F(g^{ij}\partial_i\phi\partial_j\phi) \, , \nonumber \\
\bar{\nabla_n}\phi &=& \frac{1}{N}
(\partial_t \phi-\bar{N}^ i\nabla_i \phi)=\frac{1}{N}
(\partial_t \phi-N^i\nabla_i + N\nabla^i\nu\nabla_i \phi) \, .
\nonumber %\\
\end{eqnarray}
In this section we constructed $U(1)$-invariant scalar field action in
the form which closely follows the original construction presented in
\cite{Horava:2010zj}.
In the next section  more elegant approach to the construction
of $U(1)$ invariant $F(\bar{R})$ HL gravity and the scalar
field action is given.

%%%%%%%%%%%%%%%%%%%%%%%%%%%%%%%%%%%%%%%%%%%%%
\subsection{Alternative Definition of $U(1)$ Invariant
$F(\bar{R})$ HL gravity}

To begin with we note that under the (local) $U_\Sigma (1)$ symmetry, the shift
function $N_i(\bx,t)$ and $\nu(\bx,t)$ are transformed as
\be
\label{U(1)1}
N_i(\bx,t) \to N_i(\bx,t) + N(t) \nabla_i \alpha(\bx,t) \, ,\quad
\nu(\bx,t) \to \nu(\bx,t) + \alpha(\bx,t)\, .
\ee
Therefore the combination
\be
\label{U(1)2}
{\bar N}_i(\bx,t) \equiv N_i(\bx,t) - N(t) \nabla_i \nu(\bx,t)\, ,
\ee
is invariant under the transformation of the local $U_\Sigma (1)$ symmetry.
Then if we replace $N_i$ with ${\bar N}_i$, one can always obtain the model with
the local $U_\Sigma (1)$ symmetry.

For example, the extrinsic curvature $K_{ij}$ could be replaced by
\be
\label{U(1)3}
K_{ij} \to \bar{K}_{ij} = \frac{1}{2N} \left( \partial_t g_{ij} - \nabla_i
{\bar N}_j - \nabla_j {\bar N}_i \right)
= K_{ij} + \frac{1}{2} \left( \nabla_i \nabla_j
+ \nabla_j \nabla_i \right) \nu \, .
\ee
Then it follows
\bea
\label{U(1)4}
&& S = \frac{1}{2\kappa^2} \int dt\, d^D \bx \sqrt{g} N B K_{ij}
\mathcal{G}^{ijikl} K_{kl} \nn
&& \to \frac{1}{2\kappa^2} \int dt\, d^D \bx \sqrt{g} N B \bar{K}_{ij}
\mathcal{G}^{ijikl} \bar{K}_{kl} \nn
&& = \frac{1}{2\kappa^2} \int dt\, d^D \bx \sqrt{g} \left\{ N B K_{ij}
\mathcal{G}^{ijikl} K_{kl}
+ 2 N B K_{ij} \mathcal{G}^{ijikl} \nabla_k \nabla_l \nu
+ B \nabla_i \nabla_j \nu \mathcal{G}^{ijikl} \nabla_k \nabla_l \nu
\right\}\, ,
\eea
In the same way one can deal with following term
\begin{eqnarray}
\label{U(1)5}
\nabla_\mu \left( n^\mu \nabla_\nu n^\mu - n^\nu \nabla_\nu n^\mu \right)
&=& \frac{1}{\sqrt{g} N} \left\{ \partial_t \left( \sqrt{g} K \right)
 - \partial_i \left( \sqrt{g} N^i K \right) \right\} \nonumber \\
&\to & \frac{1}{\sqrt{g} N} \left\{ \partial_t \left( \sqrt{g} \bar K \right)
 - \partial_i \left( \sqrt{g} N^i \bar K \right) \right\} \, . %\nonumber \\
\end{eqnarray}
The above analysis tells that if we define a ``curvature'' by
\be
\label{U(1)6}
\bar{R}= \bar{K}_{ij}\bar{K}^{ij}-\lambda \bar{K}^2
+ \frac{2\mu}{\sqrt{g} N} \left\{ \partial_t \left( \sqrt{g} \bar K \right)
 - \partial_i \left( \sqrt{g} N^i \bar{K} \right) \right\}
 - \mV(g_{ij})\, ,
\ee
the action
\be
\label{U(1)7}
S=\frac{1}{2\kappa^2}\int dtd^3x\sqrt{g}N F(\bar{R})\, ,
\ee
is invariant under the local $U_\Sigma (1)$ transformation (\ref{U(1)1}).
Finally, in order to obtain non-trivial symmetry we have to add to
$\bar{R}$ the expression $\mG(g_{ij})(\mA-a)$. Then the action
derived here coincides with the action (\ref{SFbarRf}).

Note that the equation obtained by the variation of $\nu$ gives a constraint
which kills the extra and problematic scalar mode appearing in the original
Ho\v{r}ava gravity.
For the action (\ref{U(1)7}), the equation has the following form:
\begin{eqnarray}
\label{U(1)12} 0 &=& \frac{N}{2}\left( \nabla_i \nabla_j + \nabla_j \nabla_i
\right) \left( \bar{K}^{ij} F'\left( \bar{R} \right) \right)
 - \lambda N \nabla^2 \left( \bar{K} F'\left( \bar{R} \right) \right)
\nonumber \\
&& - 2 \mu \nabla^2 \left( \partial_t F'\left( \bar{R} \right)
 - N^i \partial_i F'\left( \bar{R} \right) \right) \nonumber \\
&& + \frac{1}{\sqrt{g}}
\left(\frac{d}{dt}(\sqrt{g}F'(\bar{R})\mG)-
\nabla_i (N^i  \sqrt{g}F'(\bar{r})\mG)
 -\frac{N}{2}\nabla^ i\nabla_i
(\sqrt{g}F'(\bar{R})\mG)\right)=0\, . \nonumber \\
\end{eqnarray}
n the same way we can proceed with the action for  scalar field $\phi$ (\ref{Smatgen}).
Instead of using step by step procedure performed in previous
section one immediately makes the replacement
$N^i\rightarrow \hN^i=N^i- N\nabla^ i \nu$ that leads to the
action (\ref{Smattin}).
%%%%%%%%%%%%%%%%%%%%%%%%%%%%%%%%%%%%%%%%%%%%%%%%
%%%%%%%%%%%%%%%%%%%%%%%%%%%%%%%%%%%%%%%%%%%%%%%%

\section{Hamiltonian Formalism of $U(1)$ Invariant  $F(\tR)$ HL Gravity \label{fifth}}

Let us again consider an action
\begin{eqnarray}
\label{SFtRFinal2}
S_{F(\tilde{R})} &=&\frac{1}{\kappa^2}\int dt d^D\bx \left( \sqrt{g}N B(
\hK_{ij}\mG^{ijkl}\hK_{kl} -\mV(g)-A) \nonumber \right. \\
&& \left. +\sqrt{g}N F(A) -2\mu \sqrt{g}N\bar{\nabla}_n B\hK + 2\mu
\partial_i B \sqrt{g}g^{ij} \partial_j N \right) \nonumber \\
&& + \frac{1}{\kappa^2}\int dt d^D\bx \sqrt{g} B  \mG(g_{ij})(\mA-a) \, .
%\nonumber \\
\end{eqnarray}
Now we perform the Hamiltonian analysis of given action. Note that the
Hamiltonian analysis of $F(\tilde{R})$ HL gravity was performed previously in
\cite{Chaichian:2010yi,Kluson:2010xx,Chaichian:2010zn}
and we generalize these works to the
case of $U(1)$ invariant  $F(\bar{R})$ HL gravity.

 From (\ref{SFtRFinal2}) we find the conjugate momenta
\begin{eqnarray}
&& \kappa^2\pi^{ij} = \sqrt{g}B\mG^{ijkl}\hK_{kl}-\mu\sqrt{g}
\hnabla_n Bg^{ij} \, , \quad p_N\approx 0 \, , \quad p^i \approx 0 \, ,
\nonumber \\
&& \kappa^2 p_B = -2\mu\sqrt{g}\hK \, , \quad  p_A\approx 0  \, , \quad
p_\mA \approx 0 \, , \quad p_\nu=-\frac{1}{\kappa2}\sqrt{g}N\mG \, .
\nonumber %\\
\end{eqnarray}
 From these relations the primary constraints are found
\begin{equation}
\Phi_1: p_\mA(\bx)\approx 0 \, , \quad
\Phi_2: p_\nu(\bx)+\frac{1}{\kappa^2}\sqrt{g}N
\mG(\bx) \, , \quad p_A(\bx)\approx 0
\, , \quad p^i(\bx)\approx 0 \, , \quad p_N\approx 0 \ .
\end{equation}
The  Hamiltonian density is obtained in the form
\[
\mH= N\mH_T+N^i\mH_i \, ,
%\nonumber \\
\]
where
\begin{eqnarray}
\label{mHTA}
\mH_T&=&\frac{\kappa^2}{\sqrt{g}B}\pi^{ij}g_{ik}g_{il}\pi^{kl}
 -\frac{\kappa^2}{\sqrt{g}BD}\pi^2-\frac{\kappa^2\pi
p_B}{\sqrt{g}\mu D} \nonumber \\
&& + \frac{B\kappa^2}{4\mu^2\sqrt{g}D}(\lambda
D-1)p_B^2-\frac{1}{\kappa^2}\sqrt{g}B(\mV(g)-A)- \frac{1}{\kappa^2}\sqrt{g}F(A)
+ \frac{2\mu}{\kappa^2}\partial_i[\sqrt{g}g^{ij}\partial_j B]
\nonumber \\
&& - 2 \nu \nabla_i\nabla_j \pi^{ij} + \nu \nabla^i \nabla_i B
+\frac{1}{2\kappa^2}\mG(R)B\nabla^i\nabla_i \nu
\, ,  \nonumber \\
\mH_i&=&-2g_{il} \nabla_k \pi^{kl}+ p_B
\nabla_i B-\frac{1}{\kappa^2} \sqrt{g}B\mG(R)\nabla_i \nu \, ,
%\nonumber \\
\end{eqnarray}
where
\begin{equation}
\mG_{ijkl}=\frac{1}{2}(g_{ik}g_{jl}+g_{il}g_{jk})
 -\frac{\lambda}{D\lambda-1}g_{ij}g_{kl} \, .
\end{equation}
According to the general analysis of constraint systems \footnote{For
review, see \cite{Henneaux:1992ig}.} we should consider following Hamiltonian
\[
H = \int d^D\bx (N\mH_T+N^i\mH_i+ v^1\Phi_1+v^2\Phi_2+v^A p_A
+w_i p^i+w_N p_N)  - \frac{1}{\kappa^2} \int d^D\bx
\sqrt{g}\mG(R) \mA \, ,
\]
where $v^{1,2}$, $v^A$, $w_i$, and $w_N$ are Lagrange
multipliers related to corresponding primary constraints.
Now the requirement of the conservation
of the primary constraints $p_N \approx 0$,
$p_i(\bx)\approx 0$, $\Phi_{1,2}(\bx)\approx 0$ and
$p_A(\bx)\approx 0$ implies following secondary ones
\begin{eqnarray}
\label{defG}
\partial_t \Phi_1&=&\pb{\Phi_1,H}= -\frac{1}{\kappa^2} B\sqrt{g}\
\mG\equiv - \Phi_1^{II} \approx 0 \, , \nonumber \\
\partial_t p_N&=&\pb{p_N,H}= -\int d^D\bx \mH_T
\equiv -\bT\approx 0 \, , \nonumber \\
\partial_t  p_i&=&\pb{p_i,H}= -\mH_i\approx 0 \, , \nonumber \\
\partial_t p_A(\bx)&=&-\sqrt{g}B + \sqrt{g}F'(A)\equiv \sqrt{g}G_A\approx 0 \, .
%\nonumber \\
\end{eqnarray}
Finally we determine the time evolution of $\Phi_2$. Using
\[
\pb{R(\bx),\pi^{ij}(\by)}= -R^{ij}(\bx)\delta(\bx-\by)+ \nabla^i
\nabla^j \delta(\bx-\by)-g^{ij} \nabla_k \nabla^k\delta(\bx-\by)
\]
we find
\be
\label{tPhi2}
\partial_t \Phi_2= \pb{\Phi_2,H}\approx N \frac{d\mG}{dR}
\Phi_2^{II} \, ,
\ee
where the secondary constraint $\Phi_2^{II}$ takes the form
\be
\label{Phi2II}
\Phi_2^{II}= -2 R_{ij}\pi^{ij} +\frac{2}{D}R\pi -\frac{2}{D}B\nabla_k\nabla^k
(\frac{\pi}{B})+\frac{B}{\mu D} (Rp_B-(1-D)\nabla_i\nabla^i p_B) \, .
\ee
Note that in the calculation of (\ref{tPhi2}) we used the fact that
\begin{equation}
\label{pnuH}
\pb{p_\nu,H}= -N\nabla^i\mH_i + \frac{1}{\kappa^2}\nabla_i (\sqrt{g}
N^i \mG)+\frac{N}{2\kappa^2}\nabla^i \nabla_i (\sqrt{g}\mG)\sim \Phi_1^{II}
\approx 0 %\  \nonumber %\\
\end{equation}
using that the right hand of this equation is proportional
to $\Phi_1^{II}$ or its covariant derivatives
and hence it vanishes on constraint surface $\Phi_1^{II}\approx 0$.
In fact, we previously  simplified the equation (\ref{Phi2II})
where  all terms proportional to the constraint $\Phi_1^{II}$ were ignored.

Now one can summarize the results. We have
the primary constraints $p_N\approx 0$, $p_i\approx 0$, $p_\mA\approx 0$,
$p_\nu\approx 0$, $\Phi_1^{II}\approx 0$, $\Phi_2^{II}\approx 0$,
$\mH_i\approx 0$, $\bT\approx 0$.
Then the total Hamiltonian takes the form
\begin{eqnarray}
H &=& N\bT + v^Np_N \nonumber \\
&& +\int d^D\bx (N^i\mH_i+v^i p_i+ v^\mA p_\mA+v^\nu p_\nu+
v^{1}_{II}\Phi^{II}+v^{2}_{II}\Phi_2^{II}+ v^Ap_A+v^G G_A ) \, , \nonumber %\\
\end{eqnarray}
where we included an expression $\Phi_1^{II}\mA$ into the expression
$v_{II}^1 \Phi_1^{II}$.

Let us analyze the consistency of the secondary constraints
with the time development of the system for the case $F''(A)\neq 0$.
To begin with  note that one can write $\mH_i$ as
$\mH_i=-2g_{ik}\nabla_l\pi^{kl} + p_B\partial_i B
+p_\nu\partial_i\nu- \Phi_1^{II}\partial_i\nu$ so that
it is natural to consider as the smeared form of
diffeomorphism constraint the following expression
\begin{equation}
\bT_S(M^i)= \int d^D\bx M^i( -2g_{ik}\nabla_l\pi^{kl}
+p_B\partial_i B+p_\nu\partial_i\nu)
\end{equation}
that is the  generator of spatial diffeomorphism. Then
using the fact that the total Hamiltonian is manifestly
invariant under spatial diffeomorphism
one finds that $\bT_S$ is preserved on the constraint surface.
In case of the constraint $p_\nu$, using (\ref{pnuH}) and also
the fact that $\pb{p_\nu, \Phi_1^{II}}=\pb{p_\nu, \Phi_2^{II}}=0$
it follows that this constraint is
preserved during the time evolution of the system
and also that it is the first class constraint.

Let us now consider the constraint $\Phi_1^{II}$.
Its time development is governed by equation
\be
\label{parttPhi1II}
\partial_t\Phi_1^{II}= \pb{\Phi_1^{II},H_T}\approx
N\frac{d\mG}{dR}\Phi_2^{II} +\int d^D\bx v^2_{II}(\bx)
\pb{\Phi_1^{II},\Phi_2^{II}(\bx)} \, .
%\nonumber %\\
\ee
To proceed further we note that from the structure of
the constraint $\Phi_1^{II}$ and $\Phi_2^{II}$ we clearly have
\begin{equation}
\pb{\Phi_1^{II}(\bx),\Phi_2^{II}(\by)} \neq 0 \, .
\end{equation}
Say differently, $\Phi_1^{II}$ together with
$\Phi_2^{II}$ are the second class constraints.
Then the requirement that the right side of
the equation (\ref{parttPhi1II}) has to vanish implies that
the Lagrange multiplier $v^2_{II}$ has to be equal to zero.
On the other hand the time evolution of $\Phi_2^{II}$
is governed by equation
\be
\label{parttPhi2II}
\partial_t \Phi_2^{II}=N\pb{\Phi_2^{II},\bT}+\int d^D
\bx \left(v^G(\bx)\pb{\Phi_2,G(\bx)}
+ v^1_{II}(\bx)\pb{\Phi_2^{II},\Phi_1^{II}(\bx)}\right) \, .
%\nonumber %\\
\ee
Note also that $\pb{\Phi_2^{II}(\bx),G_A(\by)}\neq 0$
as follows from  (\ref{defG}) and from (\ref{Phi2II}).
On the other hand the time evolution of $G_A$ is governed by equation
\begin{equation}
\label{parttGA}
\partial_t G_A= \pb{G_A,H_T}=N\pb{G_A,\bT}+\int d^D\bx \left(
v^2_{II}(\bx)\pb{G_A,\Phi^{II}_2(\bx)}+v^A(\bx)\pb{G_A,p_A(\bx)}\right)
\end{equation}
Since $\pb{p_A(\bx), G_A(\by)}=-\sqrt{g}F''(A) \delta(\bx-\by)$
and since $v_{II}^2=0$ we see that requirement of the vanishing of
the right side of (\ref{parttGA}) determines $v^A$ as
a function of canonical variables.
On the other hand the requirement of the preservation
of the constraint $p_A$
\begin{equation}
\partial_t p_A=\pb{p_A,H_T}
\approx \int d^D\bx v^G\pb{p_A,G_A(\bx)}=0
\end{equation}
implies $v^G=0$.
However using this equation in (\ref{parttPhi2II}) we can again determine
the Lagrange multiplier $v^1_{II}$ as function of canonical variables.
At this place we see that the requirement of the preservation
of the constraints $\Phi_{1,2}^{II}$, $p_A$, $G_A$
determines the corresponding Lagrange multipliers
$v^{1,2}_{II}$, $v^A$, $v^G$ and consequently no new
constraints have to be imposed.

Now we are ready to determine the number of physical degrees of freedom.
To do this note that there are $D(D+1)$ gravity phase space variables $g_{ij}$,
$\pi^{ij}$, $2D$ variables $N_i,p^i$, $2$ variables $\mA$, $p_\mA$, $2$
variables $B$, $p_B$, $2$ variables $A$, $p_A$ and $2$ variables
$\nu,p_\nu$.
In summary the total number of degrees of freedom is
$N_\mathrm{D.o.f}=D^2+3D+8$.
On the other hand we have $D$  first class constraints $\mH_i\approx 0$, $D$
first class constraints  $p_i\approx 0 $, $2$ first class constraints
$p_\nu\approx 0$, $p_\mA\approx 0$ and four second class
constraints $\Phi_1^{II}$, $\Phi_1^{III}$, $p_A$, $G_A$.
Then there are $N_\mathrm{f.c.c.}=2D+2$ first class constraints and
$N_\mathrm{s.c.c.}=4$  second class constraints. The number of
physical degrees of freedom is \cite{Henneaux:1992ig}
\begin{equation}
\label{pdf}
N_\mathrm{D.o.f.}-2N_\mathrm{f.c.c.}-N_\mathrm{s.c.c.}= (D^2-D-2)+1
\end{equation}
that exactly corresponds to the number
of the phase space  physical degrees of freedom of $D+1$ dimensional
$F(R)$ gravity.
For example, for  $D=3$ the equation (\ref{pdf}) gives
$4$ phase space degrees of freedom corresponding
massless graviton and $2$ phase degrees of freedom corresponding
to the scalar. Note also that there is still global
Hamiltonian constraint that has to be imposed and
also that all second class constraints have to be solved.
Solving of $p_A=0$, $G_A=0$ one can express $A$ as
a function of $B$, at least in principle. Unfortunately
solving the second class constraints
$\Phi_1^{II}$, $\Phi_2^{II}$ is very difficult in the full generality.
On the other hand, it is easy to see that in
linearized approximation these constraints
can be solved as $h=0$, $\pi=0$ where
$h$ is the trace part of the metric
fluctuation and $\pi$ is its conjugate momenta.

The previous analysis was valid for the case
when $F''(A)\neq 0$. Let us now discuss
the second case when $F'(A)=1$
\footnote{Generally we could have $F'(A)=C$ for constant $C$.
However we show below  that in this case $B=C$ and consequently $C$
can be eliminated by redefinition of $\kappa$.}.
For $F'(A)=1$ we see that  the Poisson bracket between $p_A$ and
$G_A$ is zero.
Then the equation (\ref{parttGA}) implies additional constraint
(Note that $v^2_{II}=0$)
\begin{equation}
\label{parttGA2}
\partial_t G_A = \pb{G_A,H_T}=N\pb{G_A,\bT}
\equiv \frac{N \kappa^2}{\sqrt{g}\mu D}G_A^{II} \, ,
\end{equation}
where
\begin{equation}
G_A^{II}=\pi-(\lambda D-1)\frac{B p_B}{2\mu} \, .
\end{equation}
Due to the fact that $\pb{p_A,G_A}=0$ one finds that $p_A\approx 0$
is the first class constraint. On the other hand we have
following non-zero Poisson bracket
\begin{equation}
\pb{G_A(\bx),G_A^{II}(\by)}=\frac{(\lambda D-1)}{2\mu}
B(\bx)\delta(\bx-\by)
\end{equation}
that implies that $G_A$ and $G_A^{II}$ are the second class constraints.
The analysis of the remaining constraints is the same as in case
$F''(A)\neq 0$ so that we have following set of constraints.
There are $D$ first class constraints $\mH_i\approx 0$, $D$ first
class constraints  $p_i\approx 0 $, $3$ first class constraints
$p_\nu\approx p_\mA \approx 0$, $p_A\approx 0$ and four
second class constraints $\Phi_1^{II}$, $\Phi_1^{III}$, $G_A$, $G_A^{II}$.
Then we have $N_\mathrm{f.c.c.}=2D+3$ first
class constraints and $N_\mathrm{s.c.c.}=4$ second class constraints.
The first class constraint $p_A=0$ can be eliminated with the
gauge fixing condition $A=1$.
Solving the constraint $G_A=0$ we obtain $B=1$ while
solving the constraint $G_A^{II}=0$ we find
\begin{equation}
\label{pBpi}
p_B=\frac{2\mu \pi}{\lambda D-1} \, .
\end{equation}
Inserting this result into $\mH_T$ given in
(\ref{mHTA}) one gets that it takes the form
\[
\mH_T = \frac{\kappa^2}{\sqrt{g}}\left(\pi^{ij}g_{ik}g_{il}\pi^{kl}
 -\frac{\lambda}{\lambda D-1}\right)\pi^2 -\frac{1}{\kappa^2}\sqrt{g}\mV(g)
 - 2 \nu \nabla_i\nabla_j \pi^{ij} + \frac{1}{2\kappa^2}\sqrt{g}\mG(R)
\nabla^i\nabla_i \nu \, , %\nonumber %\\
\]
that corresponds to the Hamiltonian constraint
of non-relativistic covariant HL gravity whose
explicit form can be found in \cite{Kluson:2010zn}.
In the same way we insert (\ref{pBpi}) into (\ref{Phi2II}) and we find
\[
\Phi_2^{II}=-2R_{ij}\pi^{ij}+ 2\frac{\lambda}{D\lambda-1}R\pi
+ \frac{1-\lambda}{\lambda D-1}\nabla_i\nabla^i\pi
%\nonumber %\\
\]
which again coincides with the constraint found in \cite{Kluson:2010zn}.
In other words in case when $F''(A)=0$ the Hamiltonian
structure of $U(1)$ invariant $F(\bar{R})$
gravity coincides with the Hamiltonian
structure of non-relativistic covariant HL gravity.
This result can be considered as a nice confirmation of our procedure.

%%%%

%%%%%%%%%%%%%%%%%%%%%%%%%%%%%%%%%%%%%%%%%%%%%%%%%%%%%%%5
%%%%%%%%%%%%%%%%%%%%%%%%%%%%%%%%%%%%%%%%%%%%%%%%%%%

\section{Study of Fluctuations around Flat Background in $U(1)$ Invariant $F(\tR)$ HL
Gravity \label{sixth}}

Let us analyze the spectrum of fluctuations in case of $U(1)$
invariant $F(\tR)$ HL gravity for the special case $\mu=0$.
For simplicity we assume that $F(\bar{R})$ Ho\v{r}ava-Lifshitz gravity has flat
space-time as its solution with the background values of the fields
\begin{equation}
g^{(0)}_{ij}=\delta_{ij} \, , \quad N^{(0)}=1 \, , \quad N_i^{(0)}=0 \, ,
\quad  \mA^{(0)}=0 \, , \quad \nu^{(0)}=0 \, .
\end{equation}
Note that for the flat background the equation of motion for $B$ and $A$ takes
the form
\begin{equation}
\label{eqABflat}
\mV(g^{(0)})-A^{(0)}=0 \, , \quad B^{(0)}-F'(A^{(0)})=0
\end{equation}
that implies that $A^{(0)}$, $B^{(0)}$ are constants. In order to find the
spectrum of fluctuations we expand all fields  up to linear order around this
background
\begin{eqnarray}
&& g_{ij}=\delta_{ij}+\kappa h_{ij} \, , \quad
N_i=\kappa n_i \, , \quad  N=1+\kappa n \,  , \nonumber \\
&& A=A^{(0)}+\kappa a \, , \quad B=B^{(0)}+\kappa b \, , \quad
\mA=\mA^{(0)}+\kappa\tilde{\mA} \, ,
\quad \nu=\nu^{(0)}+ \kappa\tilde{\nu} \, . \nonumber %\\
\end{eqnarray}
Since $n$ depends on $t$ only, its equation of motion gives one integral
constraint. This constraint does not affect the number of local degrees of
freedom.
For that reason it is natural  to consider the equation of motion for
$h_{ij}$, $n_i$ and $\nu$ only.
We further decompose the field $h_{ij}$ and $n_i$
into their irreducible components
\begin{equation}
h_{ij}=s_{ij}+\partial_i w_j+\partial_j
w_i +\left(\partial_i\partial_j -\frac{1}{D}\delta_{ij}
\partial^2 \right)M+\frac{1}{D}\delta_{ij}h \, ,
\end{equation}
where the scalar $h=h_{ii}$ is the trace part of $h_{ij}$ while $s_{ij}$
is symmetric, traceless and transverse
\begin{equation}
\partial^i s_{ij}=0  \, , \partial^i= \delta^{ij}\partial_j \
\end{equation}
and $w_i$ is transverse
\begin{equation}
\partial^i w_i=0 \, .
\end{equation}
In the same way we decompose $n_i$
\begin{equation}
n_i=u_i+\partial_i C
\end{equation}
with $u_i$ transverse $\partial^i u_i=0$.
In what follows we fix the spatial diffeomorphism symmetry by fixing the gauge
\begin{equation}
w_i=0 \, , \quad  M=0 \, .
\end{equation}
We begin with the equation of motion for $N_i$
\begin{equation}\label{eqNi}
2\nabla_j (B\mG^{ijkl}\hK_{kl}) +B\mG(R)\nabla_i\nu=0 \, .
\end{equation}
and for $\nu$
\[
2\nabla_i\nabla_j (BN\mG^{ijkl}\hK_{kl}) + \frac{1}{\sqrt{g}}
\left(\frac{d}{dt}(\sqrt{g}B \mG)  -\nabla_i (\sqrt{g} BN^i \mG)
 -\frac{1}{2}\nabla_i\nabla^i (\sqrt{g}B N\mG)\right)=0 \, .
\]
Combining these equations and using
the equation of motion for $\mA$ one obtains
\begin{equation}
\label{eqmA}
\mG(R)=0
\end{equation}
Then
\begin{equation}
\label{nu2} B\frac{d}{dt}\mG - BN^i\nabla_i \mG
-\frac{N}{2}(B\nabla_i\nabla^i \mG + 2 \nabla_i B\nabla^i \mG) =0 \ .
\end{equation}
Further, in the linearized approximation the equation of motion
(\ref{eqNi}) takes the form
\be
\label{eqNilin}
 -2B_0(1-\lambda)\partial^i (\partial_k\partial^k C)
+B_0\partial_k\partial^k u^i+\frac{1}{D} (1-D\lambda)\partial^i
\dot{h}+ 2(1-\lambda)B_0\partial^i (\partial_j\partial^j \tilde{\nu}) =0
%\nonumber %\\
\ee
using the fact that $\mG(R_0)=0$ and also
$\partial^i s_{ij}=\partial^j s_{ij}=0$ and $\delta^{ij}s_{ji}=0$.
Let us now focus on the solution of the
constraint $\mG(R)=0$ in the linearized approximation.
Let $R^{(D)}_0$ is the solution of the
equation of motion and let us consider
the perturbation around this equation.
These perturbations have to obey the equation
\[
\mG(R_0+\delta R) = \mG(R_0)+\frac{d\mG}{dR}\delta R
= \frac{d\mG}{dR}(R_0)\delta R=0 \, .
%\nonumber %\\
\]
We see that in order to eliminate the scalar graviton we have to demand that
$\frac{d\mG}{dR}(R_0)\neq 0$.
To proceed further  note that
\begin{equation}
\delta R_{ij}= \frac{1}{2}[\nabla^{(0)}_i\nabla^{(0)k}
h_{jk}+ \nabla_j^{(0)}\nabla^{k(0)}h_{ik} -\nabla_k^{(0)}\nabla^{k(0)}h_{ij}
 -\nabla^{(0)}_i\nabla^{(0)}_j h]
\end{equation}
where $\nabla^{(0)}$ is the covariant derivative calculated using the
background metric $g_{ij}^{(0)}$.
Then in the flat background it follows
\[
\delta R = \frac{\kappa}{D}(1-D)\partial^k\partial_k h %\  \nonumber %\\
\]
so that the condition $\delta R=0$ implies $\partial_k\partial^k h=0$.
Then $h=h(t)$. However, in this case the fluctuation mode does not obey the
boundary conditions that we implicitly assumed. Explicitly we demand that all
fluctuations vanish at spatial infinity.
For that reason one should demand that $h=0$.
Then it is easy to see that the equation (\ref{nu2}) is trivially solved.
In the linearized approximation the equation of motion for $g_{ij}$ takes the form
\begin{eqnarray}
\label{eqgij}
&& \frac{1}{2}(\ddot{s}_{ij}+\frac{1}{D}\delta_{ij}(1-\lambda D) \ddot{h}-\partial_i
\dot{u}_j-2\partial_i\partial_j \dot{C}+ 2\partial^i\partial^j \tilde{\nu}-2\lambda
\delta^{ij}\partial_k\partial^k \tilde{\nu} +2\lambda \delta_{ij}\partial_k\partial^k
\dot{C}) \nonumber \\
&& - \frac{\delta \mathcal{V}_2}{ g_{ij}} +\frac{d\mG}{dR}[
\partial^i\partial^j  (\tilde{\mA}- \tilde{a}) +\delta^{ij}\partial_k\partial^k(\tilde{\mA}
 - \tilde{a})] =0
%\nonumber %\\
\end{eqnarray}
where
\begin{equation}
\tilde{a}=\frac{d\tilde{\nu}}{dt} + \frac{1}{2}\partial_k\partial^k \tilde{\nu} \, .
\end{equation}
Note that we have not fixed the $U(1)$ gauge symmetry yet. It turns out that
it is natural to fix it as
\begin{equation}
\nu=0 \, .
\end{equation}
Then the trace of the equation (\ref{eqgij}) is equal to
\be
\label{traceeqg}
(1-\lambda D)\partial_k\partial^k \dot{C} + \delta^{ij}\frac{\delta \mathcal{V}_2}
{\delta g_{ij}} -\frac{d\mG}{dR}(1-D)\partial_i\partial^i
\tilde{\mA} =0 \, .  %\nonumber %\\
\ee
Let us again consider the equation of motion (\ref{eqNilin}) and take its
$\partial^i$.
Then using the fact that  $\partial^i u_i=0$ one gets the
condition $C=f(t)$ that again with suitable boundary conditions implies
$C=0$. However then inserting this result in (\ref{eqNilin}) we find
\begin{equation}
\partial_k\partial^k u_i=0
\end{equation}
that also implies $u_i=0$.

To proceed further we now assume that $\mV_2=-R$ so that
$\frac{\delta \mV_2}{\delta g_{ij}}=-\frac{1}{2}\partial_k
\partial^k s_{ij}-\frac{D-2}{2D} (\partial_i\partial_j -\delta_{ij}\partial^k)h$.
Clearly the trace of this equation is proportional
to $h$ and hence it implies following equation for $\tilde{a}$
\begin{equation}
\partial_k\partial^k \tilde{\mA}=0 \, .
\end{equation}
Imposing again the requirement that $\tilde{\mA}$ vanishes at spatial
infinity we find that the only solution of given equation is $\tilde{\mA}=0$.
Finally the equation of motion for $g_{ij}$ gives following result
\begin{equation}
\ddot{s}_{ij}+\partial_k\partial^k s_{ij}=0 \, .
\end{equation}
In other words, it is demonstrated that under assumption that $U(1)$ invariant
$F(\bar{R})$ HL gravity has flat space-time as its
solution it follows that the perturbative spectrum contains the
transverse polarization of the graviton only.
Clearly, this result may be generalized for general version of
theory with arbitrary parameter $\mu$.

Finally we consider the linearized equations of motion for $A$ and $B$. In
case of $A$ one gets
\begin{equation}
\label{eqa}
 -b+F''(A_0)a=0
\end{equation}
while in case of $B$ we obtain
\begin{equation}
\label{eqb}
 -\frac{d\mathcal{V}}{dR}(R_0) \delta R-a=0
\end{equation}
Using the fact that $\delta R\sim h=0$ we get from (\ref{eqb}) and from (\ref{eqa})
\begin{equation}
a=b=0 \, .
\end{equation}
In other words there are no fluctuations corresponding to the scalar fields $A$ and $B$.
One can compare this situation with the conventional $F(R)$ gravity where the
mathematical equivalence of the theory with the Brans-Dicke theory implies the
existence of propagating scalar degrees of freedom.
In our case, however, the fact that $U(1)$ invariant $F(\bar{R})$
HL gravity is invariant under the foliation preserving diffeomorphism
allows us to consider theory without kinetic term for $B$ ($\mu=0$).

Thus, it seems U(1) extension of $F(R)$ HL gravity may lead to solution of
the problem of scalar graviton.

%%%%%%%%%%%%%%%%%%%%%%%%%%%%%%5
%%%%%%%
%%%%%%%

\section{Cosmological Solutions of
$U(1)$ Invariant $F(\tR)$ HL gravity \label{seventh}}

Let us investigate the FRW cosmological
solutions for the theory described by
action (\ref{U(1)7}). Spatially-flat
FRW metric is now assumed \be
\label{U(1)13}
ds^2=-N^2dt^2+a^2(t)\sum_{i=1}^3
\left(dx^{i}\right)^2\, , \ee and we
choose the gauge fixing condition for
the local $U_\Sigma (1)$ symmetry as
follows, \be \label{U(1)14} \nu = 0\, .
\ee In the FRW metric (\ref{U(1)13}),
one gets $N_i=0$ and $\bar{R}$ and
$\bar{K}_{ij}$ only depend on the
cosmological time $t$. Therefore, the
constraint equation (\ref{U(1)12}) can
be satisfied trivially. For the metric
(\ref{U(1)13}) and the gauge fixing
condition, the scalar $\bar{R}$ is
given by \be \label{U(1)15}
\bar{R}=\frac{3(1-3\lambda
+6\mu)H^2}{N^2}+\frac{6\mu}{N}\frac{d}{dt}\left(\frac{H}{N}
\right)\, . \ee The second FRW equation
can be obtained by varying the
action(\ref{U(1)7}) with respect to the
spatial metric $g_{ij}$, which yields
\begin{eqnarray}
\label{U(1)16}
0&=&F(\bar{R})-2(1-3\lambda+3\mu)\left(\dot{H}+3H^2\right)F'(\bar{R})
 -2(1- 3\lambda) \dot{\bar{R}}F''(\bar{R})+\nonumber \\
&& +2\mu\left(\dot{\bar{R}}^2F^{(3)}(\bar{R})
+\ddot{\bar{R}}F''(\bar{R})\right)+\kappa^2p_m\, .
%\nonumber \\
\end{eqnarray}
Here the contribution from the matter is included.  $p_m$ expresses
the pressure of a perfect fluid that fills the Universe, and  $N=1$.
Note that this equation becomes the usual second FRW equation for
conventional $F(R)$ gravity by setting the constants $\lambda=\mu=1$.
The variation with respect to $N$ yields the following
global constraint
\be
\label{U(1)17}
0=\int d^3x\left[F(\bar{R})-6(1-3\lambda +3\mu)H^2-6\mu\dot{H}+6\mu
H \dot{\bar{R}}F''(\bar{R})-\kappa^2\rho_m\right]\, .
\ee
By assuming the ordinary conservation equation for the matter fluid
$\dot{\rho}_m+3H(\rho_m+p_m)=0$, and integrating Eq.~(\ref{U(1)16}),
\be
\label{U(1)18}
0=F(\bar{R})-6\left[(1-3\lambda
+3\mu)H^2+\mu\dot{H}\right]F'(\bar{R})+6\mu H
\dot{\bar{R}}F''(\bar{R})-\kappa2\rho_m-\frac{C}{a^3}\, ,
\ee
where $C$ is an integration constant, taken to be zero,
according to the constraint equation (\ref{U(1)17}).
Eq.~(\ref{U(1)18}) corresponds to the first FRW equation.
Hence, starting from a given $F(\bar{R})$ function, and solving
Eqs.~(\ref{U(1)16}) and (\ref{U(1)17}), FRW cosmological solution
can be obtained.

Note that the obtained equations (\ref{U(1)16}) and (\ref{U(1)17})
are identical with the corresponding equations in the $F(R)$-gravity (for
recent review of observational aspects of such theory see
\cite{Capozziello:2009nq}) based on
the original Ho\v{r}ava gravity. Hence, $U(1)$ extension does not influence
the FRW cosmological dynamics.

Let us consider the theory which admits a de Sitter universe solution. We now
neglect the matter contribution by putting $p_m=\rho_m=0$.
Then by assuming $H=H_0$, Eq.~(\ref{U(1)18}) gives
\be
\label{HLF18}
0 = F\left( 3 \left(1 - 3 \lambda + 6 \mu \right) H_0^2 \right)
 - 6 \left(1 - 3\lambda + 3\mu\right) H_0^2 F'\left( 3 \left(1
 - 3 \lambda + 6 \mu \right) H_0^2 \right) \, ,
\ee
as long as the integration constant vanishes ($C=0$).
We now consider the following model:
\be
\label{HLF22}
F\left(\bar R\right) \propto \bar R + \beta \bar R^2
+ \gamma \bar R^3\, .
\ee
Then Eq. (\ref{HLF18}) becomes
\bea
\label{HLF23}
0 &=& H_0^2 \left\{ 1 - 3\lambda + 9\beta \left(1 - 3\lambda
+ 6 \mu \right) \left( 1 - 3\lambda + 2\mu \right) H_0^2 \right. \nn
&& \left. + 9\gamma \left(1 - 3\lambda + 6 \mu \right)^2
\left( 5 - 15\lambda + 12 \mu \right) H_0^4 \right\}\, ,
\eea
which has the following two non-trivial solutions,
\be
\label{HLF24}
H_0^2 = - \frac{ \left( 1 - 3\lambda + 2\mu
\right) \beta }{2 \left(1 - 3\lambda + 6 \mu \right) \left( 5 - 15\lambda
+ 12 \mu \right) \gamma} \left( 1 \pm \sqrt{
1 - \frac{4 \left(1 - 3\lambda \right)\left( 5 - 15\lambda + 12 \mu
\right) \gamma} { 9 \left( 1 - 3\lambda
+ 2\mu \right)^2 \beta^2} } \right)\, ,
\ee
as long as the r.h.s. is real and positive. If
\be
\label{HLF25}
\left| \frac{4 \left(1 - 3\lambda \right)\left( 5 - 15\lambda + 12 \mu
\right) \gamma} { 9 \left( 1 - 3\lambda + 2\mu \right)2 \beta^2} \right|
\ll 1\, ,
\ee
one of the two solutions is much smaller than the other solution.
Then one may regard that the larger solution corresponds to the inflation
in the early universe and the smaller one to the late-time acceleration.

More examples of $F\left(\bar R\right)$ theory which can contain more
than one dS solution, such that inflation and dark energy epochs can be
explained under the same mechanism
Ref.~\cite{NojiriEtAl} may be considered. First of all, as
generalization of the model (\ref{HLF22}), a general polynomial
function may be discussed
\be
F\left(\bar R\right)=\sum_{n=1}^m \alpha_n\bar R^n\, ,
\label{D1}
\ee
Here ${\alpha_n}$ are coupling constants. Using the equation
(\ref{HLF18}), it yields the algebraic equation,
\be
0=\sum_{n=1}^m \alpha_n\bar R^n_0-2\frac{1-3\lambda+3\mu}{1-3\lambda+6\mu}\bar
R_0\sum_{n=1}^m n\alpha_n\bar R^{n-1}_0\, .
\label{D2}
\ee
By a qualitative analysis, one can see that the number of positive real roots,
i.e., of the de Sitter points, depends completely on the sign of the coupling
constants ${\alpha_n}$.
Then, by a proper choice, $F\left(\bar R\right)$
gravity can well explain dark energy and inflationary epochs in a unified
natural way. Even it could predict the existence of more than two accelerated
epochs, which could resolve the coincidence problem.

Let us now consider an explicit example
\be
F\left(\bar R\right)=\frac{\bar R}{\bar R(\alpha \bar
R^{n-1}+\beta)+\gamma}\, ,
\label{D3}
\ee
where ${\alpha, \beta, \gamma, n}$ are  constants. By introducing this
function in (\ref{HLF18}), it is straightforward to show that for the
function (\ref{D3}), there are several de Sitter solutions. In order to
simplify this example, let us consider the case  $n=2$, where the equation
(\ref{HLF18}) yields,
\be
\gamma-3\gamma\lambda-3\beta
H_0^2(1-3\lambda+6\mu)^2+27\alpha
H_0^4(-1+3\lambda-4\mu)(1-3\lambda+6\mu)^2=0
\, .
\label{D4}
\ee
The solutions are given by
\bea
&& H_0^2 = \frac{1}{18\alpha(1-3\lambda+6\mu)^2(-1+3\lambda-4\mu)}
\left\{ \beta(1-3\lambda+6\mu)^2 \right. \nn
&& \left. \quad \pm\sqrt{(1-3\lambda+6\mu)^2
\left[12\alpha\gamma(-1+3\lambda)(-1+3\lambda-4\mu)
+\beta^2(1-3\lambda+6\mu)^2\right]} \right\} \, .
\label{D5}
\eea
Then, by a proper choice of the free
parameters of the model, two positive roots of the equation
(\ref{D4}) are solutions. Hence, such a model can explain
inflationary and dark energy epochs in unified manner.

\section{Discussion \label{eighth}}

In summary, in this work we aimed to resolve (at least, partially) the
inconsistency problems of the projectable HL gravity.
First of all, it is demonstrated that some versions of $F(R)$ HL gravity may
have stable de Sitter solution and unstable flat space solution.
As a result, the spectrum analysis showing the presence of scalar graviton
is not applied. The whole spectrum analysis should be redone for de Sitter
background.

Second, $U(1)$ extension of $F(R)$ HL gravity is formulated in two
alternative approaches. Hamiltonian structure of $U(1)$ invariant
$F(\bar{R})$ gravity is investigated in all detail.
The whole constraints system is derived and different
particular cases corresponding to conditions for derivatives of function
$F$ are studied. It is demonstrated that in some cases the Hamiltonian
structure of the theory coincides with the one of $U(1)$ invariant HL gravity
that conforms consistency of our approach.
The analysis of fluctuations of $U(1)$ invariant $F(\bar{R})$ HL
gravity is performed.
It is shown that like in case of $U(1)$ HL gravity the
scalar graviton ghost does not emerge.
This opens good perspectives for consistency of such class of models.
It is also interesting that spatially-flat FRW equations for $U(1)$ invariant
$F(\bar{R})$ gravity turn out to be just the same as for the one without
$U(1)$ symmetry. This indicates that all (spatially-flat FRW) cosmological
predictions of viable conventional $F(R)$ gravity are just the same as for
its HL counterpart (with special parameters choice).

\section*{Acknowledgements}

This research  was supported in part by the Czech Ministry of Education under
Contract No. MSM 0021622409 (JK), by MICINN (Spain) project FIS2006-02842
and AGAUR (Catalonia) 2009SGR-994 (SDO) and by Global COE Program of Nagoya
University (G07) provided by Ministry of Education, Culture, Sports, Science
\&  Technology and by the JSPS Grant-in-Aid for Scientific Research
(S) ·22224003(SN).
%\vskip 5mm

%%%%%%%%%%%%%%%%%%%%%%%%%%%%%%%%%%%%%%
%%%%%%% Thebibligraphy %%%%%%%%%%%
%%%%%%%%%%
%%%%%%%%%%%%%%%%%%%%%%%%%%%%%%%%%%%%%


\begin{thebibliography}{20}

%\cite{Horava:2009uw}
\bibitem{Horava:2009uw}
P.~Horava,
%\emph{``Quantum Gravity at a Lifshitz Point,''}
Phys.\ Rev.\  D {\bf 79} (2009) 084008
[arXiv:0901.3775 [hep-th]];
JHEP {\bf 0903} (2009) 020
[arXiv:0812.4287 [hep-th]];
arXiv:0811.2217 [hep-th].
%%CITATION = PHRVA,D79,084008;%%

%\cite{Padilla:2010ge}
\bibitem{Padilla:2010ge}
A.~Padilla,
%  \emph{``The good, the bad and    the ugly .... of Horava gravity,''}
arXiv:1009.4074 [hep-th]; \\
S.~Mukohyama, arXiv:1007.5199 [hep-th]; \\
T.~P.~Sotiriou, arXiv:1010.3218 [hep-th].
%%CITATION = ARXIV:1009.4074;%%

\bibitem{gravitation}
C.~W.~Misner, K.~S.~Thorne,
J.~A.~Wheeler, \textit{Gravitation},
W. H. Freeman and Company, 1973,
San Francisco; \\
R.~L.~Arnowitt, S.~Deser and
C.~W.~Misner, arXiv:gr-qc/0405109.
%%CITATION = GR-QC/0405109;%%

%\cite{Rubakov:2008nh}
\bibitem{Rubakov:2008nh}
V.~A.~Rubakov and P.~G.~Tinyakov,
%``Infrared-modified gravities and massive gravitons,''
Phys.\ Usp.\  {\bf 51} (2008) 759
[arXiv:0802.4379 [hep-th]].
%%CITATION = PHUSE,51,759;%%

%\cite{Sotiriou:2009bx}
\bibitem{Sotiriou:2009bx}
T.~P.~Sotiriou, M.~Visser and S.~Weinfurtner,
%\emph{``Quantum gravity without Lorentz invariance,''}
JHEP {\bf 0910} (2009) 033
[arXiv:0905.2798 [hep-th]].
%%CITATION = JHEPA,0910,033;%%

%\cite{Huang:2010rq}
\bibitem{Huang:2010rq}
Y.~Huang, A.~Wang and Q.~Wu,
%\emph{``Stability of the de Sitter
%spacetime in Horava-Lifshitz theory,''}
Mod.\ Phys.\ Lett.\  A {\bf 25} (2010) 2267
[arXiv:1003.2003 [hep-th]]; \\
A.~Wang and Q.~Wu,
%\emph{``Stability of spin-0 graviton and strong coupling in Horava-Lifshitz
%theory of gravity,''}
arXiv:1009.0268 [hep-th].
%%CITATION = MPLAE,A25,2267;%%

%\cite{Blas:2009yd}
\bibitem{Blas:2009yd}
D.~Blas, O.~Pujolas and S.~Sibiryakov,
% \emph{``On the Extra Mode and
% Inconsistency of Horava Gravity,''}
JHEP {\bf 0910} (2009) 029
[arXiv:0906.3046 [hep-th]];
arXiv:0909.3525 [hep-th];
arXiv:0912.0550 [hep-th]; \\
M.~Li and Y.~Pang,
JHEP {\bf 0908} (2009) 015
[arXiv:0905.2751 [hep-th]]; \\
M.~Henneaux, A.~Kleinschmidt and G.~L.~Gomez,
arXiv:0912.0399 [hep-th]; \\
J.~Bellorin and A.~Restuccia,
arXiv:1010.5531 [hep-th];
arXiv:1004.0055 [hep-th]; \\
A.~Kobakhidze,
Phys.\ Rev.\  D {\bf 82} (2010) 064011
[arXiv:0906.5401 [hep-th]]; \\
J.~M.~Pons and P.~Talavera,
Phys.\ Rev.\  D {\bf 82} (2010) 044011
[arXiv:1003.3811 [gr-qc]].
%%CITATION = PHRVA,D82,044011;%%

%\cite{Blas:2010hb}
\bibitem{Blas:2010hb}
D.~Blas, O.~Pujolas and S.~Sibiryakov,
%\emph{``Models of non-relativistic
% quantum gravity: the good, the bad and the
%healthy,''}
arXiv:1007.3503 [hep-th].
%%CITATION = ARXIV:1007.3503;%%

%\cite{Kluson:2010xx}
\bibitem{Kluson:2010xx}
J.~Kluson,
%%\emph{``Note About Hamiltonian
%%Formalism of Modified $F(R)$
%%Ho\v{r}ava-Lifshitz Gravities and Their
%%Healthy Extension,''}
Phys.\ Rev.\  D {\bf 82} (2010) 044004
[arXiv:1002.4859 [hep-th]];
JHEP {\bf 1007} (2010) 038
[arXiv:1004.3428 [hep-th]].
%  %%CITATION = PHRVA,D82,044004;%%

%\cite{Horava:2010zj}
\bibitem{Horava:2010zj}
P.~Horava and C.~M.~Melby-Thompson,
%\emph{``General Covariance in Quantum Gravity at a Lifshitz Point,''}
Phys.\ Rev.\  D {\bf 82} (2010) 064027
[arXiv:1007.2410 [hep-th]].
%%CITATION = PHRVA,D82,064027;%%

%\cite{Greenwald:2010fp}
\bibitem{Greenwald:2010fp}
J.~Greenwald, V.~H.~Satheeshkumar and A.~Wang,
%``Black holes, compact objects and solar system tests in non-relativistic
%general covariant theory of gravity,''
arXiv:1010.3794 [hep-th]; \\
A.~Wang and Y.~Wu,
arXiv:1009.2089 [hep-th]; \\
J.~Alexandre and P.~Pasipoularides,
arXiv:1010.3634 [hep-th].
%%CITATION = ARXIV:1010.3794;%%

%\cite{Huang:2010ay}
\bibitem{Huang:2010ay}
Y.~Huang and A.~Wang,
%\emph{``Nonrelativistic general
%covariant theory of gravity with a
%running constant $\lambda$,''}
arXiv:1011.0739 [hep-th].
%%CITATION = ARXIV:1011.0739;%%

%\cite{daSilva:2010bm}
\bibitem{daSilva:2010bm}
A.~M.~da Silva,
%\emph{``An Alternative Approach for General Covariant Horava-Lifshitz
%Gravity and
%  Matter Coupling,''}
arXiv:1009.4885 [hep-th].
%%CITATION = ARXIV:1009.4885;%%

%\cite{Kluson:2010zn}
\bibitem{Kluson:2010zn}
J.~Kluson,
%``Hamiltonian Analysis of Non-Relativistic Covariant RFDiff Horava-Lifshitz
%Gravity,''
arXiv:1011.1857 [hep-th].
%%CITATION = ARXIV:1011.1857;%%

%\cite{Kluson:2010na}
\bibitem{Kluson:2010na}
J.~Kluson,
%``Horava-Lifshitz Gravity And Ghost Condensation,''
arXiv:1008.5297 [hep-th].
%%CITATION = ARXIV:1008.5297;%%

%\cite{Chaichian:2010yi}
\bibitem{Chaichian:2010yi}
M.~Chaichian, S.~Nojiri, S.~D.~Odintsov, M.~Oksanen and A.~Tureanu,
%\emph{``Modified F(R) Horava-Lifshitz gravity: a way to accelerating FRW
%cosmology,''}
Class.\ Quant.\ Grav.\  {\bf 27} (2010)
185021 [arXiv:1001.4102 [hep-th]]; \\
S.~Carloni, M.~Chaichian, S.~Nojiri, S.~D.~Odintsov, M.~Oksanen and
A.~Tureanu,
Phys.\ Rev.\  D {\bf 82}, 065020 (2010)
[arXiv:1003.3925 [hep-th]].
%%CITATION = CQGRD,27,185021;%%

%\cite{Nojiri:2010wj}
\bibitem{Nojiri:2010wj}
S.~Nojiri and S.~D.~Odintsov,
%\emph{``Unified cosmic history in modified gravity: from F(R) theory to
%Lorentz
%  non-invariant models,''}
arXiv:1011.0544 [gr-qc].
%%CITATION = ARXIV:1011.0544;%%

%\cite{Chaichian:2010zn}
\bibitem{Chaichian:2010zn}
M.~Chaichian, M.~Oksanen and A.~Tureanu,
%\emph{``Hamiltonian analysis of non-projectable modified F(R)
%Ho\v{r}ava-Lifshitz
%  gravity,''}
Phys.\ Lett.\  B {\bf 693} (2010) 404
[arXiv:1006.3235 [hep-th]].
%%CITATION = PHLTA,B693,404;%%

%\cite{Kluson:2009xx}
\bibitem{Kluson:2009xx}
J.~Kluson,
%\emph{``New Models of f(R) Theories of Gravity,''}
Phys.\ Rev.\  D {\bf 81} (2010) 064028
[arXiv:0910.5852 [hep-th]]; \\
J.~Kluson,
JHEP {\bf 0911} (2009) 078
[arXiv:0907.3566 [hep-th]].

%\cite{Capozziello:2009nq}
\bibitem{Capozziello:2009nq}
S.~Capozziello, M.~De Laurentis and V.~Faraoni,
%\emph{``A bird's eye view of
%f(R)-gravity,''}
arXiv:0909.4672 [gr-qc].
%%CITATION = ARXIV:0909.4672;%%

%\cite{Nojiri:2006ri}
\bibitem{Nojiri:2006ri}
S.~Nojiri and S.~D.~Odintsov,
%\emph{``Introduction to modified
%gravity and gravitational alternative
%for dark energy,''}
eConf {\bf C0602061} (2006) 06
[Int.\ J.\ Geom.\ Meth.\ Mod.\ Phys.\  {\bf 4} (2007) 115]
[arXiv:hep-th/0601213].
%%CITATION = 00436,4,115;%%

%\cite{Buchbinder:1992rb}
\bibitem{Buchbinder:1992rb}
I.~L.~Buchbinder, S.~D.~Odintsov and I.~L.~Shapiro,
\emph{``Effective action in quantum
gravity,''}
%\href{/spires/find/hep/www?irn=2762668}{SPIRES entry}
{\it  Bristol, UK: IOP (1992) 413 p}

\bibitem{FRdeSitter}
G.~Cognola, E.~Elizalde, S.~D.~Odintsov, P.~Tretyakov and S.~Zerbini,
%\emph{``Initial and final
% de Sitter universes from modified $f(R)$ gravity,''}
Phys.\ Rev.\  D {\bf 79} (2009) 044001
[arXiv:0810.4989 [gr-qc]].
%%CITATION = PHRVA,D79,044001;%%

%\cite{Kiritsis:2009sh}
\bibitem{Kiritsis:2009sh}
E.~Kiritsis and G.~Kofinas,
%\emph{``Horava-Lifshitz Cosmology,''}
%  Nucl.\ Phys.\  B {\bf 821}, 467 (2009)
[arXiv:0904.1334 [hep-th]]; \\
D.~Capasso and A.~P.~Polychronakos,
JHEP {\bf 1002} (2010) 068
[arXiv:0909.5405 [hep-th]]; \\
T.~Suyama,
arXiv:0909.4833 [hep-th]; \\
J.~M.~Romero,
V.~Cuesta, J.~A.~Garcia and J.~D.~Vergara,
Phys.\ Rev.\  D {\bf 81}, 065013 (2010)
[arXiv:0909.3540 [hep-th]]; \\
A.~E.~Mosaffa,
arXiv:1001.0490 [hep-th]; \\
E.~Kiritsis and G.~Kofinas,
JHEP\ {\bf 1001}, 122 (2010)
[arXiv:0910.5487 [hep-th]]; \\
S.~K.~Rama,
arXiv:0910.0411 [hep-th].
%%CITATION = ARXIV:0910.0411;%%
%%CITATION = NUPHA,B821,467;%%

%\cite{Henneaux:1992ig}
\bibitem{Henneaux:1992ig}
M.~Henneaux and C.~Teitelboim,
\emph{``Quantization of gauge systems,''}
%\href{http://www.slac.stanford.edu/spires/find/hep/www?irn=2824396}{SPIRES
%entry}
{\it  Princeton, USA: Univ. Pr. (1992) 520 p}; \\
J.~Govaerts,
\emph{"
Hamiltonian Quantization And Constrained Dynamics,''}
%\href{http://www.slac.stanford.edu/spires/find/hep/www?irn=2864207}{SPIRES
%entry}
{\it  Leuven, Belgium: Univ. Pr. (1991)
371 p. (Leuven notes in mathematical
and theoretical physics, B4)}.

\bibitem{NojiriEtAl}
%\cite{Elizalde:2010ep}
%\bibitem{Elizalde:2010ep}
E.~Elizalde, S.~Nojiri, S.~D.~Odintsov and D.~Saez-Gomez,
%``Unifying inflation with dark energy in modified F(R) Horava-Lifshitz
%gravity,''
arXiv:1006.3387 [hep-th].
%%CITATION = ARXIV:1006.3387;%%

\end{thebibliography}
\end{document}